\documentclass[]{aa}  

\usepackage{graphicx}
\usepackage{txfonts}
\usepackage{xcolor,colortbl}
\usepackage[colorlinks=true,citecolor=cyan,linkcolor=cyan]{hyperref}
\usepackage{multirow}
\usepackage[normalem]{ulem}
\makeatletter
\renewcommand*\aa@pageof{, page \thepage{} of \pageref*{LastPage}}
\makeatother
\newcommand{\rmaxzmax}{$R_{max}$ vs. $z_{max}$ }
\newcommand{\rmax}{$R_{max}$\ }
\newcommand{\zmax}{$z_{max}$\ }

\begin{document} 

   \title{Orbital analysis of stars in the nuclear stellar disc of the Milky Way}

   \author{N. Nieuwmunster
            \inst{1,2}
            \and
            M. Schultheis
            \inst{1}
            \and
            M. Sormani
            \inst{3}
            \and
            F. Fragkoudi
            \inst{4}
            \and
            F. Nogueras-Lara
            \inst{5}
            \and
            R. Schödel
            \inst{6}
            \and
            P. McMillan
            \inst{2,7}
            }

   \institute{Université Côte d’Azur, Observatoire de la Côte d’Azur, Laboratoire Lagrange, CNRS, Blvd de l’Observatoire, 06304 Nice, France\\
    \email{niels.nieuwmunster@oca.eu}
    \and   
    Division of Astrophysics, Department of Physics, Lund University, Box 43, SE-22100 Lund, Sweden
    \and
    Department of Physics, University of Surrey, Guildford GU2 7XH, UK
    \and
    Institute for Computational Cosmology, Department of Physics, Durham University, South Road, Durham DH1 3LE, UK
    \and
    Max-Planck Institute for Astronomy, Königstuhl 17, 69117 Heidelberg, Germany
    \and
    Instituto de Astrofísica de Andalucía (CSIC), Glorieta de la Astronomía s/n, 18008 Granada, Spain
    \and
    School of Physics \& Astronomy, University of Leicester, University Road, Leicester, LE1 7RH, UK
    }

   \date{Received December 18, 2023; accepted February 21, 2024}

 
  \abstract{
While orbital analysis studies were so far mainly focused on the Galactic halo, it is possible now to do these studies in the heavily obscured region close to the Galactic Centre.}
{We aim to do a detailed orbital analysis of stars located in the nuclear stellar disc (NSD) of the Milky Way allowing us to trace the dynamical history of this structure.}
{We integrated orbits of the observed stars in a non-axisymmetric potential. We used a Fourier transform to estimate the orbital frequencies. We compared two orbital classifications, one made by eye and the other with an algorithm, in order to identify the main orbital families. We also compared the Lyapunov and the frequency drift techniques to estimate the chaoticity of the orbits.}
{We identified several orbital families as chaotic, $z$-tube, $x$-tube, banana, fish, saucer, pretzel, 5:4, and 5:6 orbits. As expected for stars located in a NSD, the large majority of orbits are identified as $z$-tubes (or as a sub-family of $z$-tubes). Since the latter are parented by $x_{2}$ orbits, this result supports the contribution of the bar (in which $x_{2}$ orbits are dominant in the inner region) in the formation of the NSD. Moreover, most of the chaotic orbits are found to be contaminants from the bar or bulge which would confirm the predicted contamination from the most recent NSD models.}
{Based on a detailed orbital analysis, we were able to classify orbits into various families, most of which are parented by $x_{2}$-type orbits, which are dominant in the inner part of the bar.} 


   \maketitle

\section{Introduction}

The nuclear stellar disc (NSD) is a dense stellar structure in the centre of the Milky Way  and surrounds the massive nuclear star cluster (NSC) with its central massive black hole  (\citealt{Launhardt2002}). The NSD is a flattened disc with a radius of $\rm \sim 200\,pc$ and a scale height of $\rm \sim 50\,pc$ (\citealt{Launhardt2002}; \citealt{nishiyama13}, \citealt{Gallego-Cano2020}). Increasing evidence is reported that the NSD is a distinct structure from the NSC and the nuclear bulge: \citet{nogueras2020} determined the SFH in the NSD using the GALACTICNUCLEUS  data (\citealt{Nogueras2018}) and analysing the luminosity function together with stellar evolutionary models. They found that $\rm \sim 80\%$ of the stars formed more than 8\,Gyr ago, followed by a quenching phase and then by a recent star formation activity, about 1\,Gyr ago, in which about 5\% of the NSD mass was formed. While most of studies agree that the NSD has a relatively early formation time, the detailed SFH is still under discussion (see e.g. \citealt{Nogueras2023}, \citealt{Sanders2023}), and much more work is clearly needed.

 By using a large sample of KMOS observations in the NSD (\citealt{Fritz2021}), \citet{Schultheis2021} found a difference in the chemistry, that is, in the metallicity distribution function, between the NSC, NSD, and the nuclear bulge that reinforces a different formation scenario of the NSD. Furthermore, they found some evidence that metal-rich stars may have formed in the central molecular zone, while metal-poor stars show more similarities to the surrounding Galactic bulge.

Kinematic  studies relying on radial velocity measurements or proper motion studies show  evidence that the NSD is rotating (see e.g. \citealt{Lindquist1992},  \citealt{schoenrich2015}, \citealt{Fritz2021}, \citealt{Shazamanian2022}). Linking the rotation to the chemistry, \citet{Schultheis2021} found that metal-rich stars rotate faster than metal-poor stars, with some hints of counter-rotation for the most metal-poor stars.

Extragalactic studies showed that many barred galaxies host nuclear discs or rings \citep{Gadotti2019,Gadotti2020}. So far, the most likely formation scenario of a nuclear disc, called inside-out formation, is linked to the galactic bar \citep{Bittner2020}. According to this scenario, a nuclear disc is a built up from a series of gaseous rings (i.e. nuclear rings) that grow in radius over time. The growth is caused by the gas that is moved towards the galactic centre by the bar.

Based on the 3D velocities, \citet{Sormani2022} constructed axisymmetric  self-consistent equilibrium dynamical models of the NSD providing the full 6D distribution function (position and velocity) of the NSD. These models 
 provide the best description of the rotation curve  in the innermost few hundred parsecs of the Milky Way, and they are implemented in the \textit{AGAMA} (\citealt{agama}) software package. 

A powerful method for obtaining a complete picture of the properties of the individual orbits is the so-called frequency analysis (\citealt{Laskar1993}, \citealt{Valluri1998}, \citealt{vasiliev2013}) where the three fundamental frequencies of the  orbit oscillation can be extracted accurately. This frequency analysis can be used  to distinguish between regular and  chaotic orbits and to classify the main orbital families. So far, this technique has mainly been used for studies in the Galactic halo and disc. \citet{amarante2020} used frequency maps to show that the proposed wedges in the $\rm R_{apo}-z_{max}$ plane  identified by \citet{haywood2018} as possible signs of accretion can be explained by the existence of different orbital families.
\citet{koppelman2021} revealed the prominent presence of resonances. According to this, $\sim 30\%$ of the halo stars are associated with resonant families. 

In this paper, we calculate the orbital parameters of a representative sample of stars belonging to the NSD and we use frequency analysis to classify the different types of orbits present in the NSD. 

\section{Observations}
\subsection{Data sample}
We used the NSD data obtained with the KMOS \citep{KMOS} spectrometer at the ESO VLT. The detailed survey strategy and data reduction procedures are described by \citet{Fritz2021}. As well as their radial velocities, we used the proper motions derived from the VIRAC2 (Smith et al. in prep) photometric and astrometric reduction of the VVV data (\citealt{minniti2010}). These proper motions were rescaled to the Gaia absolute reference frame (\citealt{Sanders2019}). We also refer here to \citet{Sormani2022}. Our sample includes only stars that are primary sources of the survey leaving us a total of 2501 stars. 

\begin{figure*}[ht!]
   \centering
   \includegraphics[width=\hsize]{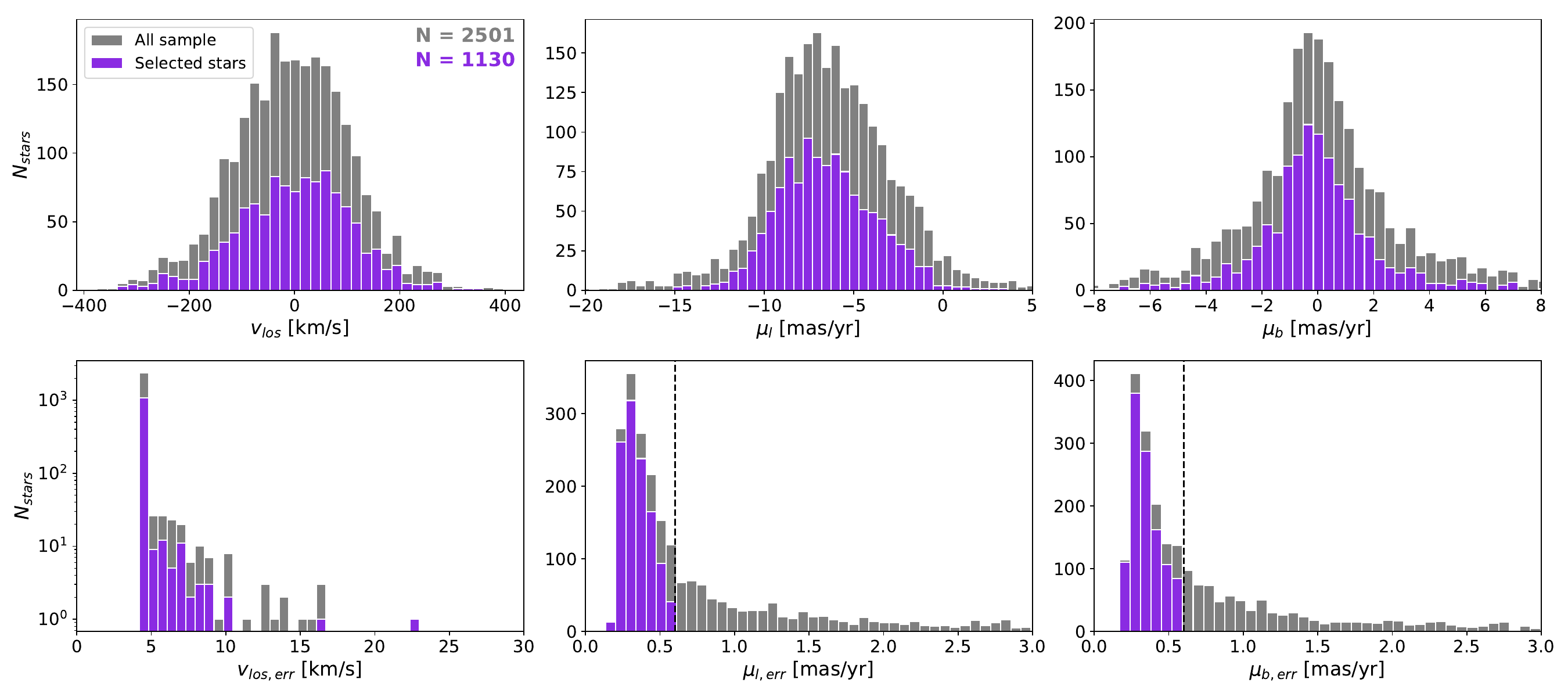}
      \caption{Upper panel: Histograms of the radial velocities and proper motions in $\rm \mu_{l}$ and $\rm \mu_{b}$, respectively. Lower panel: Same, but for the uncertainties. The grey sample is the full sample, and our final sample, used for the analysis, is shown in magenta.}
         \label{fig:hist_params}
\end{figure*}

\subsection{Catalogue selection}
\citet{Sormani2022} pointed out that stars belonging to the Galactic bar contribute significantly in the outermost fields of the NSD sample of \citet{Fritz2021}. In their Tab.~10, they quantified the contamination of the bar: It ranges from 20--30\% in the inner fields to up to 75\% in the outermost fields. For this reason, we decided to use only the innermost fields, that is, $\rm |l| < 1.5^{o}$ and $\rm |b| < 0.3^{o}$, where the probability of NSD membership is higher than 70\% (see Tab. 2 of \citealt{Sormani2022}). In addition, we used the same colour cut $\rm (H-Ks) > max(1.3,-0.0233\ K + 1.63)$ as in \citet{Schultheis2021} to remove foreground stars.

\begin{figure}[ht!]
   \centering
   \includegraphics[width=\hsize]{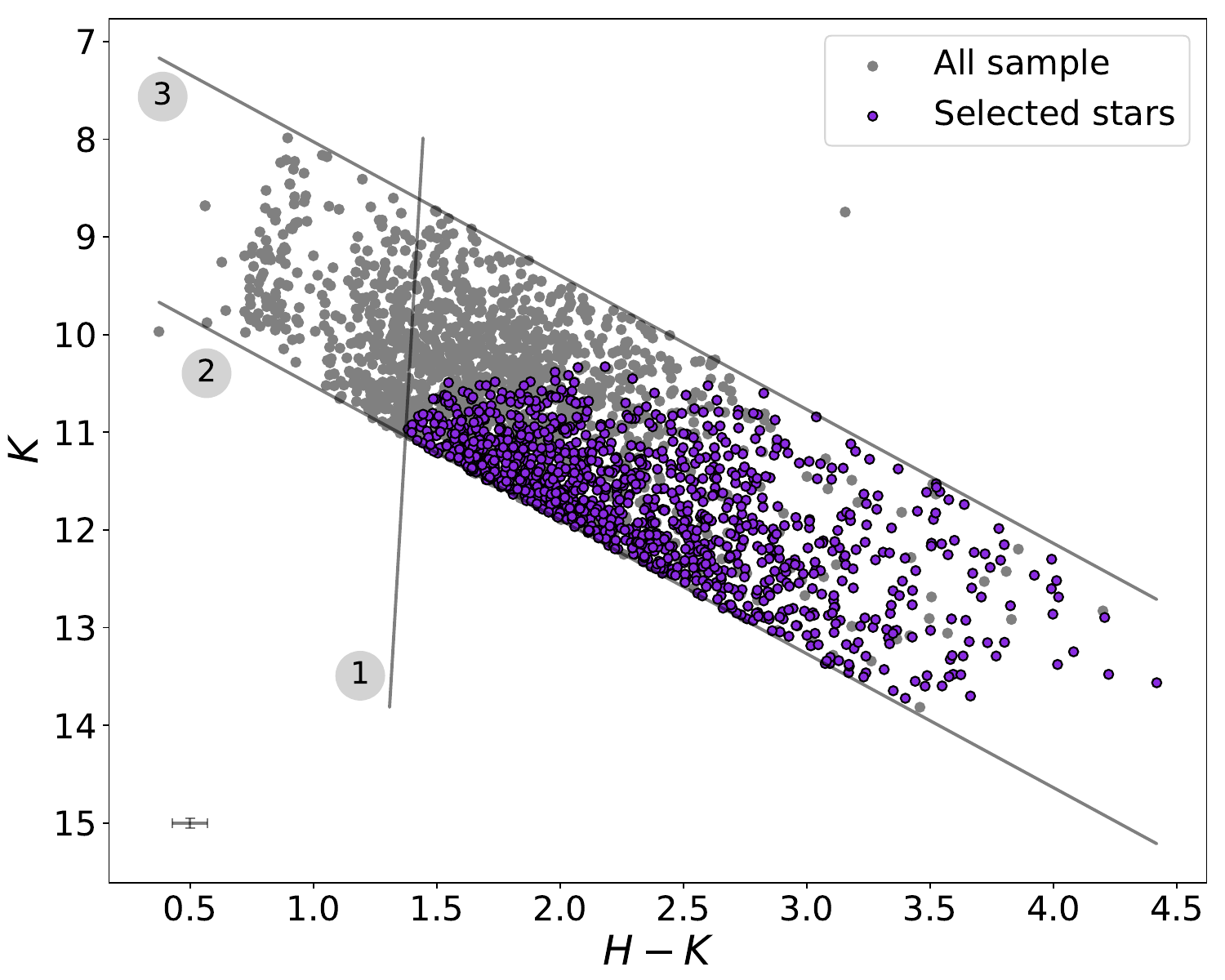}
      \caption{K vs. H--K colour magnitude diagram of the total \citet{Fritz2021} sample in grey. Our sample after application of the proper motion cuts is shown in magenta. Colour cut 1: $(H-K) > max(1.3, -0.0233 K + 1.63)$. Colour cuts 2 and 3: $6.6575<K-1.37(H-K)<9.1575$. The typical error for each axis is indicated in the lower left corner of the figure.}
         \label{fig:magplot}
\end{figure}

In order to obtain highly reliable orbital parameters, we constructed a golden sample and chose to keep stars with small proper motion uncertainties: $\rm \mu_{l,err} < 0.6\,mas\,yr^{-1}$ and $\rm \mu_{b,err} < 0.6\,mas\,yr^{-1}$ (corresponding to the $98^{th}$ percentile; see the vertical dashed line in Fig.~\ref{fig:hist_params}), leading to a total sample of 1130 stars. As pointed out by \citet{Sormani2022}, this proper motion cut also removes very bright stars ($\rm K < 10$), for which the proper motion errors become large due to saturation effects. Figure \ref{fig:magplot} shows the corresponding $\rm HK$ colour-magnitude diagram after removing the bright stars. The final sample contains stars in the interval $\rm 6.6575 < K - 1.37\ (H-K) < 9.1575$, corresponding to a dereddened magnitude  of $\rm  7.0 < K_{0} < 9.5$. The main purpose of our selection criteria was to obtain an unbiased sample in the metallicity distribution as discussed in \citet{Schultheis2021}. We caution that our sample is far from complete, which affects the fraction of the different orbital families (see Section 5). To overcome this, we plan to address this issue in a forthcoming paper by using N-body simulations of the NSD.

\section{Analysis}
\subsection{Orbit integration} \label{orbits}

We used the software package \textit{AGAMA} \citep{agama} to determine the orbital parameters. It is both fast in terms of computation time and provides methods for the easy handling of the computation of different potentials. This gives us the flexibility to test different potentials (e.g. axisymmetric, non-axisymmetric, or different bar pattern speeds).

One main source of uncertainties when calculating the orbital parameters is the distance uncertainty. While previous studies of the GC used a constant distance (e.g. 8.25\,kpc \citealt{Gravity}), we implemented a distance distribution in our analysis. As the NSD is an extended stellar feature with a scale length of $\sim$ 100\,pc (e.g. \citealt{Gallego-Cano2020}, \citealt{Sormani2022}, \citealt{Nogueras22}), we assumed this typical scale length and integrated the orbits 100 times by using different distances chosen from a Gaussian distribution ($\mu=8.2$ kpc, $\sigma=50$ pc) between $8.1$ kpc and $8.3$ kpc  (\citealt{Launhardt2002}, \citealt{Nogueras22}).\\
In addition, we carried out two other tests in which we assigned the distances of the stars in  two different ways: (i) We used the colour cuts made by \cite{Nogueras2023} to identify, the stars in our sample that belong to the close edge (i.e. bluer in H--K) and those that belong to the inner region of the NSD (i.e. redder in H--K). Typical distances of $8.05$ kpc and $8.2$ kpc with a standard deviation of $50$ pc were used for the closest edge and the inner population, respectively. (ii) We computed the most likely distances for each star, by using their probability to belonging to the NSD. The probabilities were derived from the distribution function provided by the self-consistent model from \cite{Sormani2022}. The results obtained with these different distance values are discussed in Sections \ref{sec:results} and \ref{sec:discussion}.

For the purpose of this paper, we constructed a non-axisymmetric potential by combining the potentials that correspond to the main components of the Galaxy that affect the dynamics of stars in the NSD: the inner bulge/bar, NSD, and NSC. The total density is $\rm \rho_{tot} = \rho_{bar}+ \rho_{NSD}+ \rho_{NSC}$ where (i) the bar/bulge density is taken from \cite{Launhardt2002};
(ii) We adopted the NSD density from \cite{Sormani2020} (see their Equation 27); 
(iii) We used the NSC density from \cite{Chatzopoulos2015} (see their Equation 17). We show a comparison with other potentials  in Appendix \ref{sec:potentialscomparison}.\\
We took into account a typical (clock-wise) rotation pattern speed of $\Omega_{b}= 40$\,km s$^{-1}$ kpc$^{-1}$ \citep{Portail17} for the bar component of the potential which is at an angle of $\alpha = 25^{\circ}$ from the line of sight towards the Galactic centre (GC).    \\
For the purpose of this study, it is not necessary to set a long integration time because stars located near the GC rotate quickly around it. As explained in \cite{Valluri2016}, an accurate determination of the fundamental orbital frequencies requires that orbits are integrated for at least 20 orbital periods. During $500$ Myr, most of the stars of our sample have therefore made hundreds to thousands of periods, which is enough to estimate orbital frequencies (see Section \ref{sec:freq_determination}). We chose a short timestep of $4000$ yr for a good sampling of the orbits.

\subsection{Orbital frequency determination}
\label{sec:freq_determination}
Bounded regular orbits in a triaxial potential have three fundamental frequencies, $\rm \Omega$, that determine the periodic behaviour of motion (\citealt{binneytremaine2008}). This motion can be decomposed as a Fourier sum, where the Fourier frequencies are linear combinations of the fundamental frequencies. These fundamental frequencies can be  recovered, as shown by \cite{Laskar1993}, using the so-called numerical approximation of fundamental frequencies (NAFF), which has been applied in planetary dynamics. \citet{superfreq} developed an open-source code called {\textit{SuperFreq}}, which is a Python implementation similar to the NAFF code. This code finds the fundamental frequencies (starting with the highest amplitude) by computing the Fourier spectra for the phase-space coordinates used to describe the orbit. A plot of these fundamental frequencies of a set of orbits gives then a frequency map that allows identifying resonances. Resonant orbits are those for which the fundamental frequencies $\rm \vec{\Omega} = \Omega_{x},\Omega_{y}, \Omega_{z}$ (or $\Omega_{R},\Omega_{\Phi}, \Omega_{z}$ depending on the selected coordinate system) can be measured and where $\rm \vec{n} \cdot \vec{\Omega} = 0$ for $\rm n= (n_{x},n_{y},n_{z})$, where the vector n only contains integer numbers and  at most one zero (\citealt{koppelman2021}).
For example, in Cartesian coordinates, $\rm n_{x}\Omega_{x}+n_{y}\Omega_{y}+n_{z}\Omega_{z}=0$, and the resonance is then called $n_{x}:n_{y}:n_{z}$. Moreover, boxlets are special cases of resonant orbits in which one of the integers $n_{x},n_{y},n_{z}$ is zero.\\
Frequency maps are a powerful tool for obtaining an automatic classification of the different orbital families with a much clearer separation when plotting the fundamental frequencies in Cartesian coordinates (\citealt{Valluri2016}).\\
We can classify orbits into two main categories: tubes (short-axis or long-axis)\footnote{Tube orbits circulate around a certain axis: $z$ for short-axis tubes and $x$ for long-axis tubes.}, and box orbits (chaotic orbits correspond to orbits that cannot be identified with one of these categories, see Section \ref{sec:chaos}). Based on the frequencies in Cartesian coordinates ($\Omega_{x},\Omega_{y}, \Omega_{z}$), we can identify most of the orbital families corresponding to the different resonances. However, since the chosen coordinate system only allows us to trace symmetries within the system, short-axis and long-axis tube orbits in the Cartesian case are clustered to a trivial resonance (e.g. 1:-1:0 resonance for short-axis tube orbits). It is then more effective to study orbital frequencies in several coordinate systems even though the Cartesian frequencies give us a general picture of the orbital families. To study tube orbits more precisely and identify their true resonances, it can be better to examine the frequencies in cylindrical coordinates ($\Omega_{R},\Omega_{\Phi}, \Omega_{z}$). However, \cite{Valluri2016} showed that Cartesian orbital frequencies are more reliable for bar orbit classification.\\

\begin{figure}[ht!]
   \centering
   \includegraphics[width=\hsize]{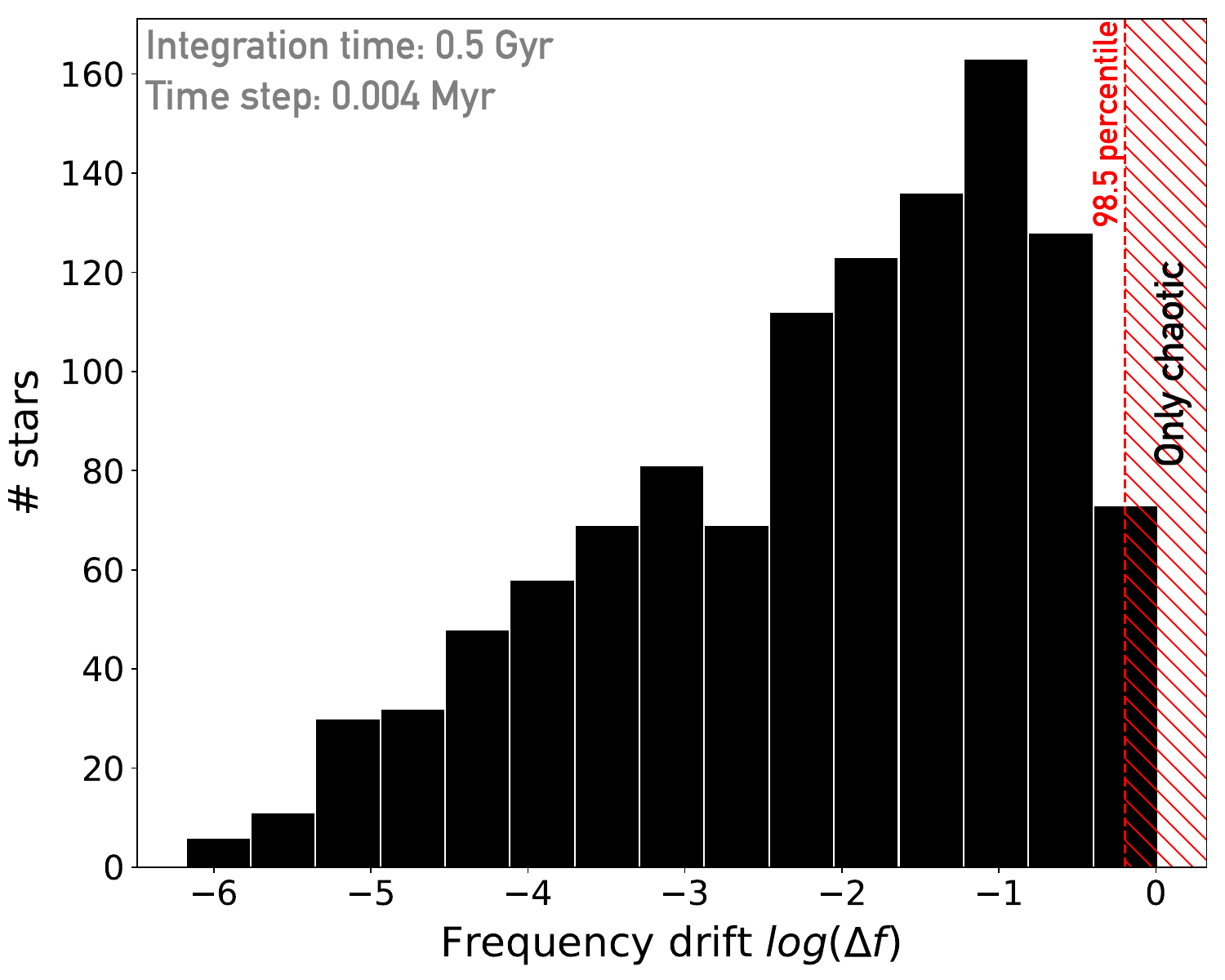}
      \caption{Histogram of the frequency drift. The dashed red area denotes the 98.5 percentile limit above which all orbits are classified as chaotic orbits.}
         \label{fig:frequency drift}
\end{figure}

\begin{figure}[ht!]
   \centering
   \includegraphics[width=\hsize]{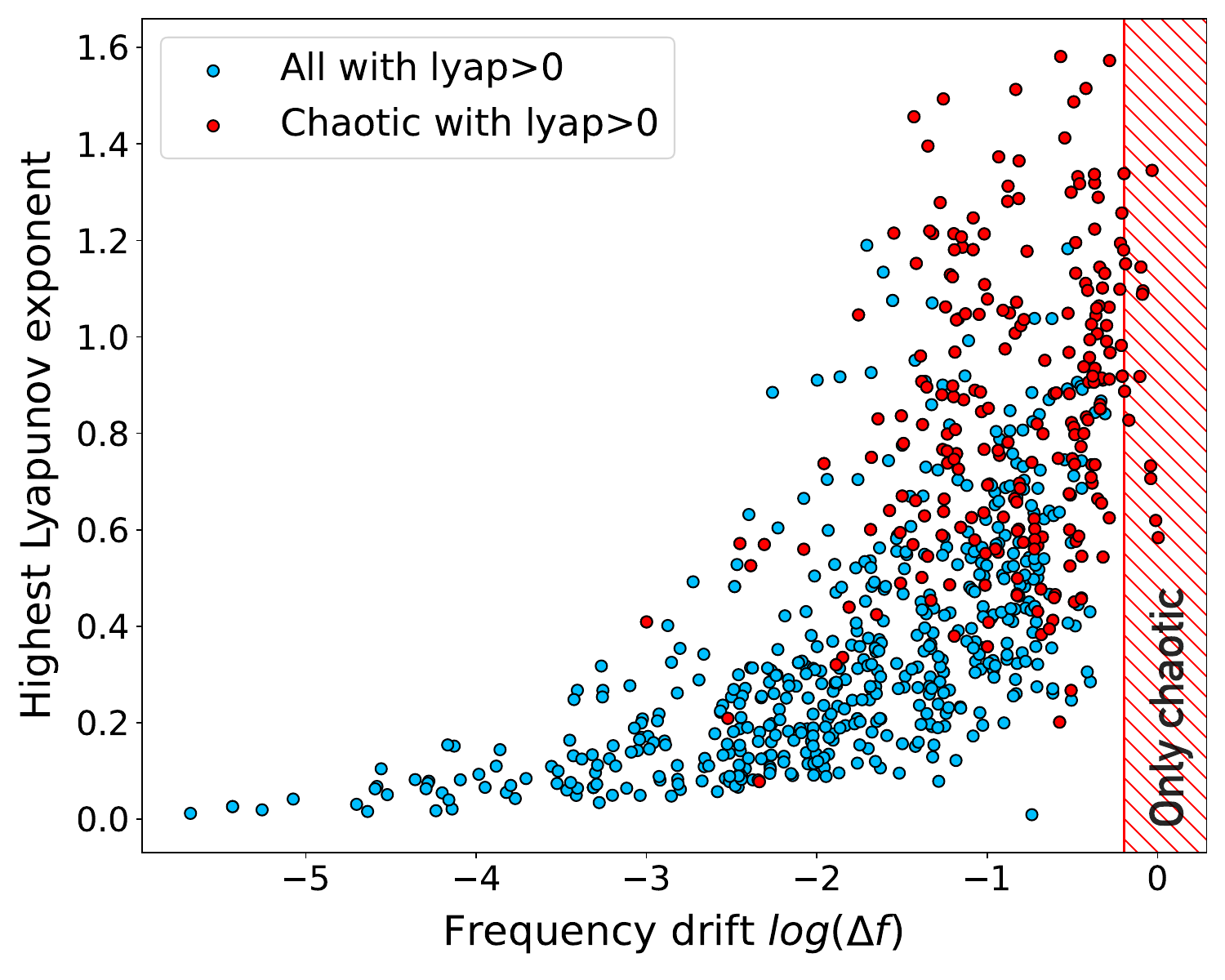}
      \caption{Comparison between highest Lyapunov exponent and frequency drift. The red points show the chaotic orbits identified with the visual classification (see Sect. \ref{sec:visual}) while the red shaded area indicates the zone containing only chaotic orbits according to the frequency drift method} (see Sect. \ref{sec:freqdrift}). Only orbits with a highest Lyapunov exponent greater than 0 are shown in this diagram.
          \label{fig:freqdrift vs lyap}
\end{figure}

\subsection{Chaoticity}
\label{sec:chaos}
To study chaoticity and particularly identify chaotic orbits, we used two different methods that we introduce in this section.
\subsubsection{Lyapunov exponent} \label{sec:lyapunov}
The Lyapunov exponent \citep{Lyapunov:1992} is a fundamental concept in the field of non-linear dynamics and chaos theory. It quantifies the sensitivity of a dynamical system to initial conditions. In a global context, the Lyapunov exponent characterizes how the trajectories in a system diverge or converge as time progresses, providing insights into the long-term behaviour of complex systems.

When considering a dynamical system, even tiny differences in initial conditions can lead to drastically different trajectories over time. The Lyapunov exponent captures this phenomenon by measuring the rate of exponential separation between initially close trajectories. A high Lyapunov exponent indicates chaotic behaviour, where trajectories diverge exponentially over time, making long-term predictions inherently uncertain. Conversely, a low Lyapunov exponent indicates convergence towards a stable equilibrium or periodic behaviour. \\
We used the highest Lyapunov exponent estimation method provided by \textit{AGAMA} \citep{agama}. For more details about the method, see Section 4.3 of \cite{vasiliev2013}.

\begin{figure}[ht!]
   \centering
   \includegraphics[width=\hsize]{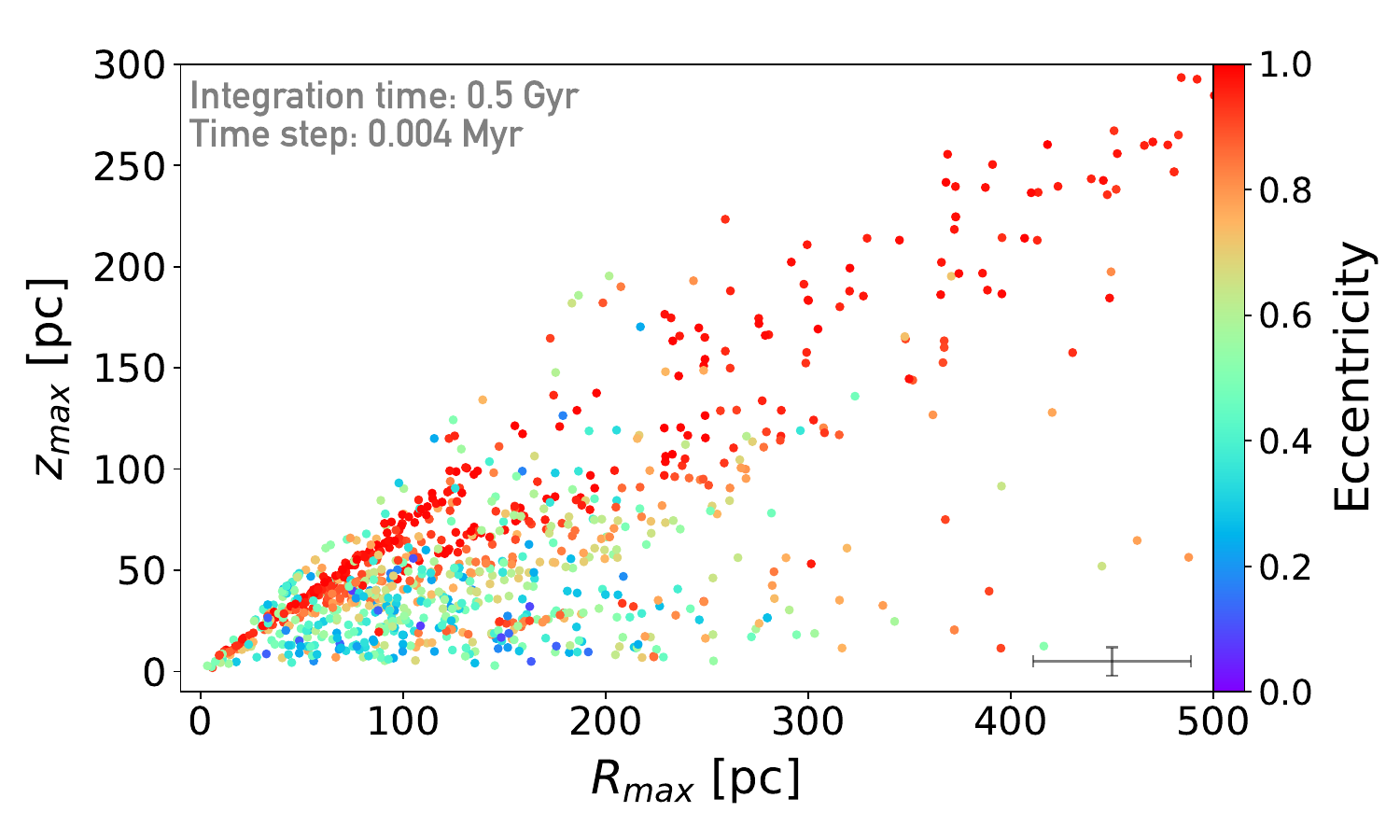}
      \caption{$R_{max}$ vs. $z_{max}$ diagram colour-coded with eccentricity. The typical error on both parameters arising from the propagation of the observational uncertainties and the distance uncertainty is indicated in the lower right corner.}
         \label{fig:rmax_zmax_ecc}
\end{figure}

\begin{figure}[ht!]
   \centering
   \includegraphics[width=\hsize]{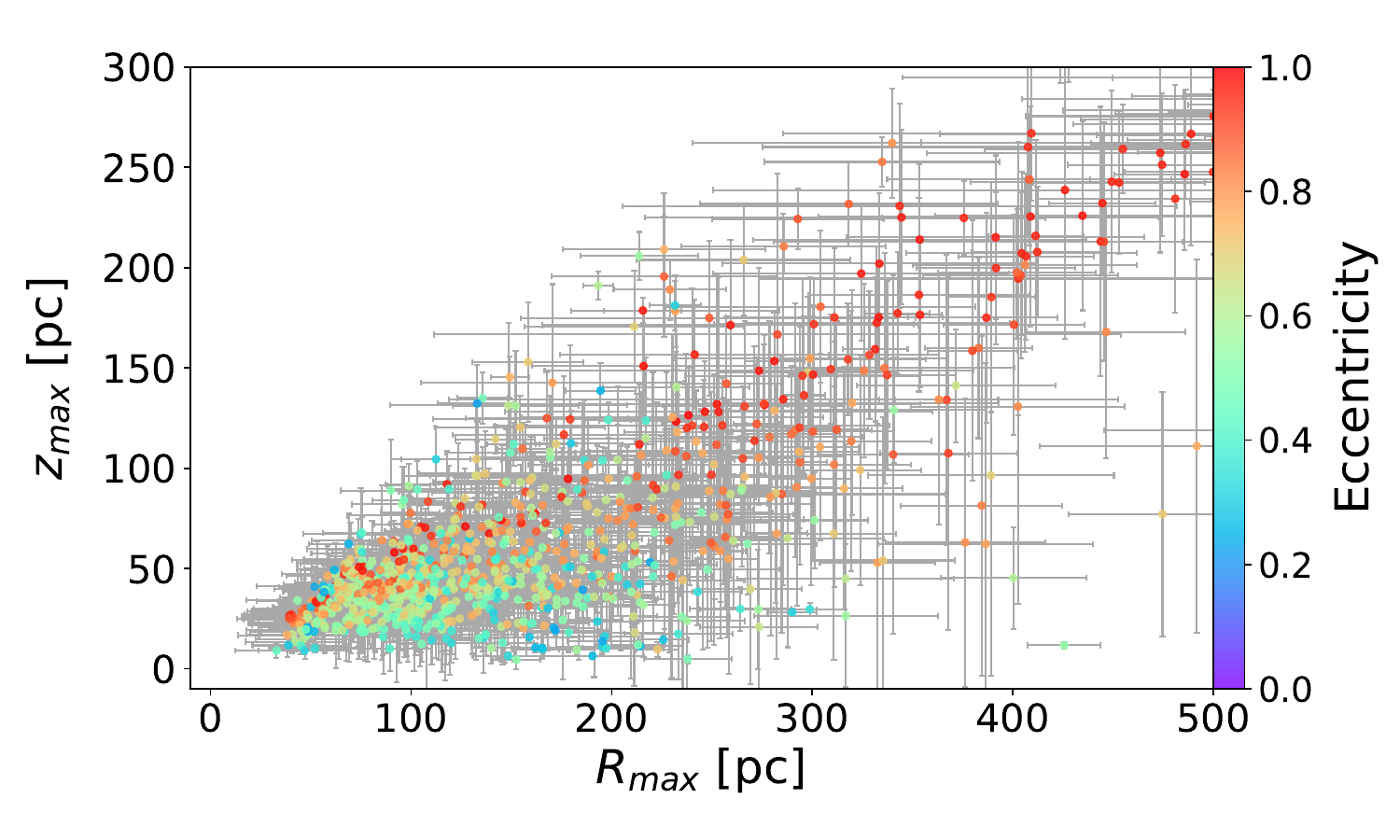}
      \caption{Average $R_{max}$ vs. $z_{max}$ diagram (colour-coded with eccentricity) by using 100 MCMC distances. The standard deviation for each star is indicated with the error bars.}
         \label{fig:rmax_zmax_mcmc_distances}
\end{figure}

\begin{figure}[ht!]
   \centering
   \includegraphics[width=\hsize]{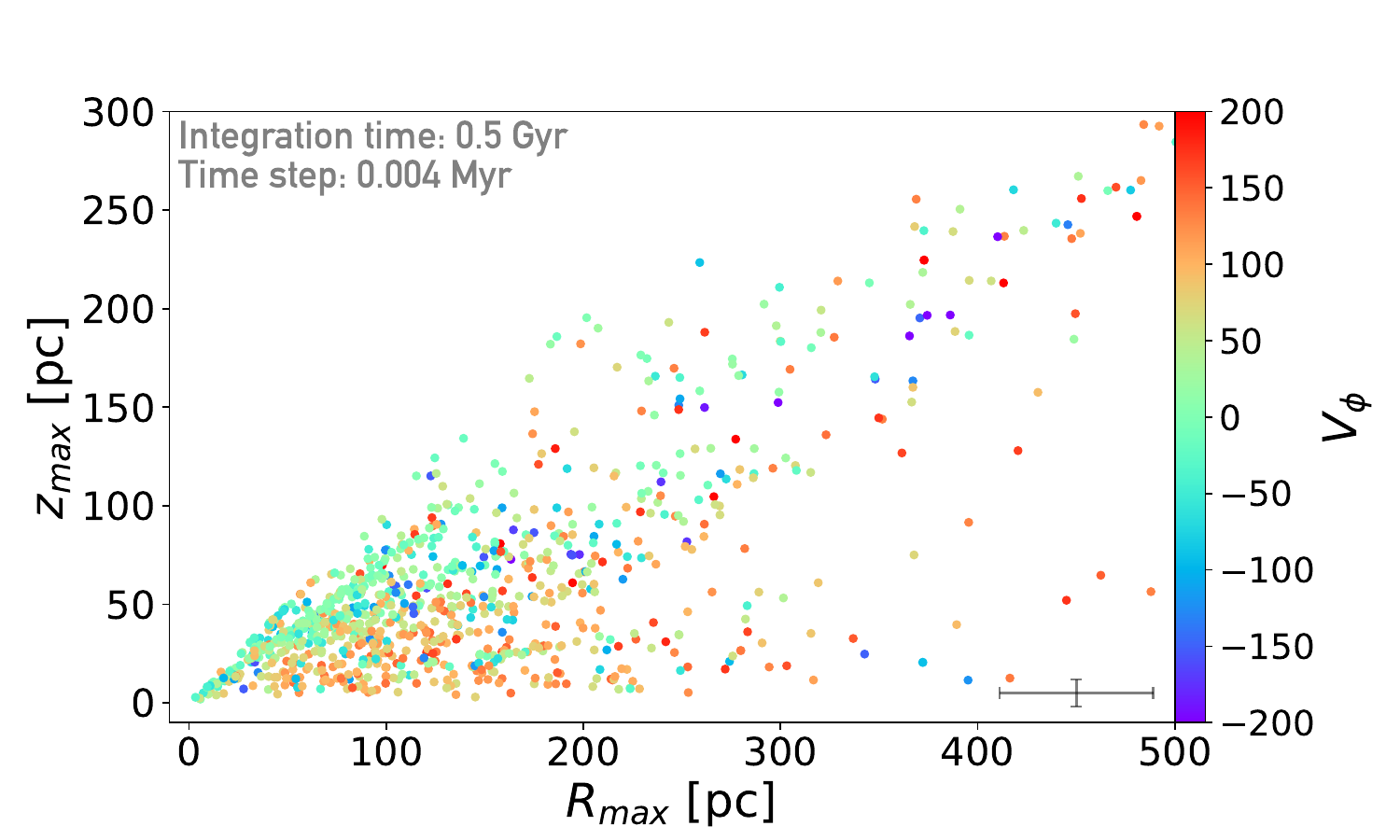}
      \caption{$R_{max}$ vs. $z_{max}$ diagram colour-coded with the rotational velocity. $V_{\phi} >0$ means clockwise (i.e. the same direction as the bar). The typical error on both parameters arising from the propagation of the observational uncertainties and the distance uncertainty is indicated in the lower right corner.}
         \label{fig:rmax_zmax_vphi}
\end{figure}

\subsubsection{Frequency drift} \label{sec:freqdrift}
Another way to study chaoticity is to directly use the orbital frequencies. \cite{valluri2010} (see their Section 3.1) showed that it is possible to measure the stochasticity of an orbit based on the change in the fundamental frequencies over two consecutive time intervals. For each frequency component $f_{i}$, they computed what they called the frequency drift:
\begin{equation}
    \log(\Delta f_{i})=\log\Bigg|\frac{\Omega_{i}(t_{1})-\Omega_{i}(t_{2})}{\Omega_{i}(t_{1})}\Bigg|
\end{equation}
where i defines the frequency component in Cartesian coordinates (i.e. $\rm \log(\Delta f_{x})$, $\rm \log(\Delta f_{y})$ and $\rm \log(\Delta f_{z})$). The highest value of the three frequency drift parameters $\rm \log(\Delta f_{i})$ is then associated with the frequency drift parameter $\rm \log(\Delta f)$. The higher the value of $\rm \log(\Delta f)$, the more chaotic the orbit. However, as shown by \citet{valluri2010},  the accuracy of the frequency analysis requires at least 20 oscillation periods in  order to avoid a misclassification of the orbits as chaotic. Figure \ref{fig:frequency drift} shows  the distribution of the frequency drift parameter $\rm \log(\Delta f)$.  In order to define a threshold value of $\rm \log(\Delta f)$ at which orbits are classified as chaotic, we  followed a similar approach as in \citet{valluri2010}. 1.5 \% of the orbits have $\rm \log(\Delta f)  > -0.2$ which we consider as the threshold of being chaotic orbits. In total, we obtained $18$ chaotic orbits in our sample. Figure ~\ref{fig:freqdrift vs lyap} shows the comparison between the frequency drift parameter and the highest Lyapunov exponent. These two measurements of the chaoticity are generally correlated but the dispersion is high. Our visual classification of chaotic orbits (see Sect.~\ref{sec:visual}, red points in Fig.~\ref{fig:freqdrift vs lyap}) nicely shows that chaotic orbits can indeed be identified based on their large Lyapunov exponent and high frequency drift parameter.

\begin{figure*}[ht!]
   \centering
   \includegraphics[width=0.9\textwidth]{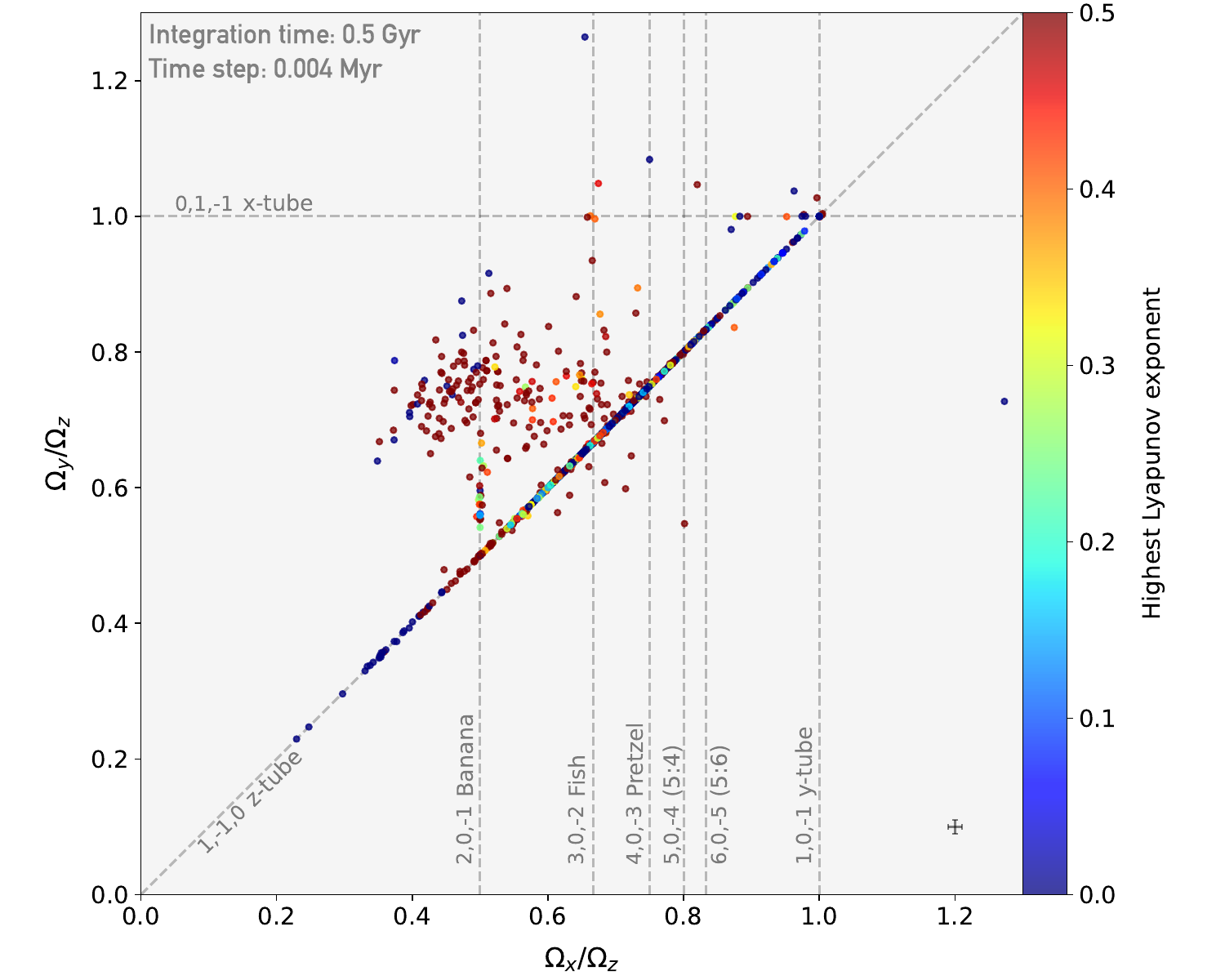}
      \caption{Frequency map in Cartesian coordinates vs. highest Lyapunov exponent. For reasons of legibility, the banana, fish, pretzel, 5:4 and 5:6 resonances are only shown here for the $(x,z)$ plane case. The typical error on both frequency ratios arising from the propagation of the observational uncertainties and the distance uncertainty is indicated in the lower right corner.}
         \label{fig:freq_map_cart}
\end{figure*}

\begin{figure*}[ht!]
   \centering
   \includegraphics[width=0.9\textwidth]{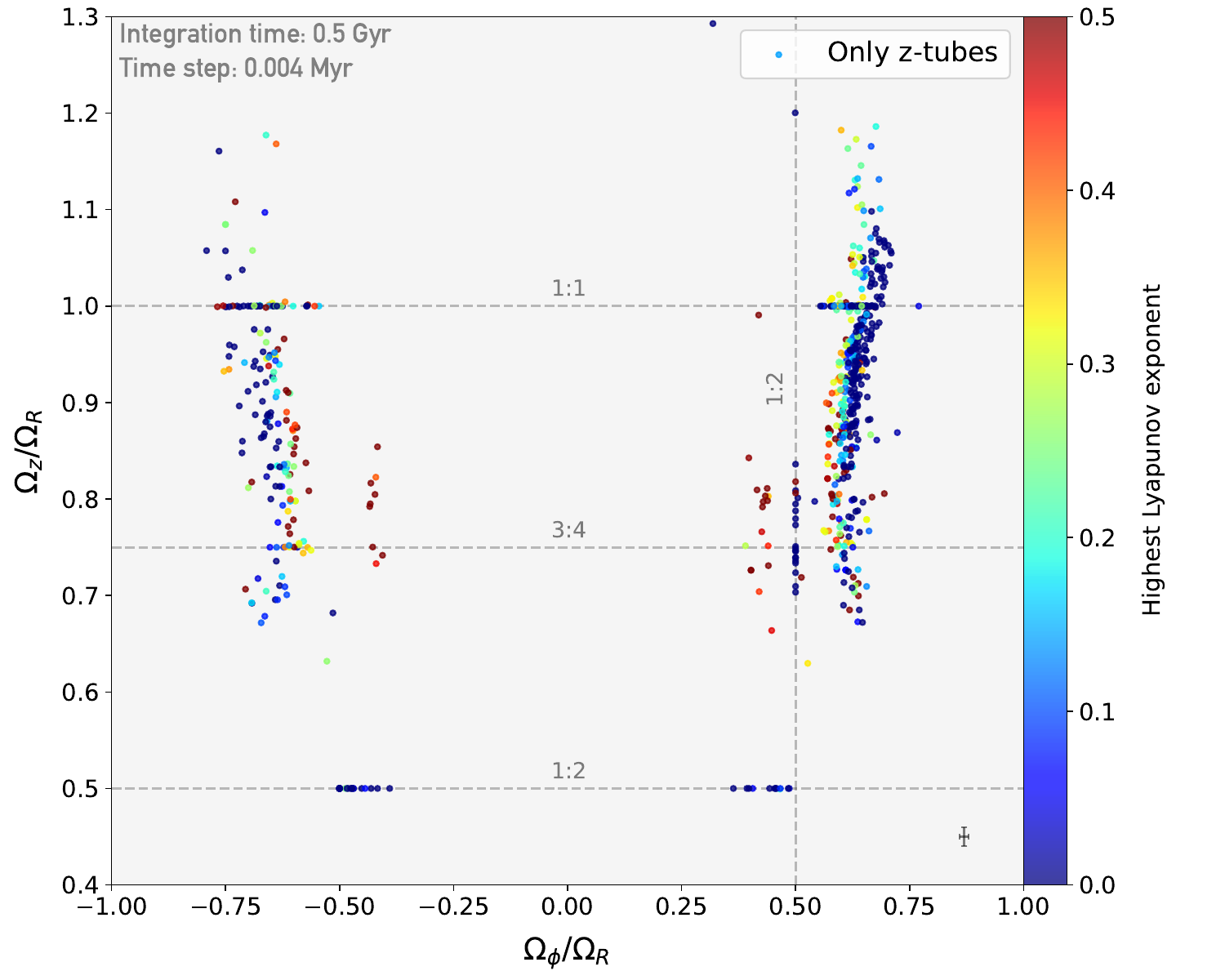}
      \caption{Frequency map in cylindrical coordinates for $z$-tubes alone (orbits for which a circulation around the $z$-axis has been detected), i.e. orbits from the resonance (1:-1:0) in Cartesian coordinates (see Fig.\ref{fig:freq_map_cart}). The horizontal and vertical lines correspond to resonances between $\Omega_{z}$ and $\Omega_{R}$ and between $\Omega_{\phi}$ and $\Omega_{R}$ respectively. The typical error on both frequency ratios arising from the propagation of the observational uncertainties and the distance uncertainty is indicated in the lower right corner.}
         \label{fig:freq_map_cyl}
\end{figure*}

\section{Results} \label{sec:results}

\subsection{\rmaxzmax diagram}

Figure~\ref{fig:rmax_zmax_ecc} shows the apocentric radius ($\rm R_{max}$) versus the maximum height ($\rm z_{max}$) of our NSD sample colour-coded by the eccentricity.

This diagram shows several features: (i) One striking feature is that the stars are not homogeneously distributed, but congregate in distinct diagonal wedges in which $\rm z_{max}$ increases with $\rm R_{max}$. (ii) Highly eccentric orbits are confined to one wedge. (iii) Some of these highly eccentric orbits extend beyond the typical radius of the NSD and are likely stars related to the Galactic bar.
Figure \ref{fig:rmax_zmax_mcmc_distances} shows the impact for a relaxed assumption of a constant distance and when the distances of the stars are instead allowed to vary within a spread of 100 pc for the NSD, using MCMC simulations (as detailed in Section 3). The key characteristics mentioned earlier clearly persist. This indicates that the uncertainty in the distances of NSD stars can safely be disregarded.

Figure~\ref{fig:rmax_zmax_vphi} shows the same feature, but as a function of rotational velocity $V_{\phi}$. The NSD clearly rotates with typical velocities of $\rm 80\,km/s$, which agrees with the works of \citet{Lindquist1992},  \citet{schoenrich2015}, \citet{Schultheis2021}, \citet{Shazamanian2022}, and \citet{Sormani2022}. The figure also shows stars with slower rotation and even counter-rotating stars, which have been identified previously by \citet{Schultheis2021}.

\subsection{Orbit classification}
We discuss and compare the two methods we used to classify orbits in various orbital families.
\subsubsection{Automatic method}\label{sec:automatic}

We call this method "automatic" because no visual inspection of the orbits is used to classify them (compared to the other classification method presented in Section \ref{sec:visual}).\\
Firstly, to differentiate tube orbits from chaotic/box orbits, we determined whether the orbit circulates around a specific axis (e.g. a circulation around the x-axis results in an $x$-tube) or not (box orbit or chaotic orbit). To do this, we verified whether there was a change in the sign of the angular momentum about an axis. We find 34\% box orbits and 66\% tube orbits among our 1130 stars. 
Secondly, to identify the different orbital families, we studied the resonances on the frequency map as explained in Section \ref{sec:freq_determination}. In the Cartesian version of the frequency map, Fig.\ref{fig:freq_map_cart}, we observe two main structures: (i) a diagonal line called the 1:-1:0 resonance, where short-axis tubes (called here after "$z$-tubes") lie, and (ii) a cloud located above the latter resonance and below the 0:1:-1 resonance that is thought to contain mainly box and chaotic orbits. In addition to these two large structures that clearly dominate our frequency map, a few orbits lie close to the 0:1:-1 resonance where long-axis tubes, called here after $x$-tubes, are located. An example of the morphology of an $x$-tube is given in Fig.\ref{fig:x_tube}. \\
Based on the three main types of orbits (box/chaotic, $z$-tube, $x$-tube), we proceeded with the classification in order to distinguish the resonant orbits that populate the two large structures of the map.\\
Classical tube orbits are simply orbits with a resonance 1:1 between two of the three frequency ratios. In the same way, we obtained banana orbits with a 2:1 resonance (see Fig.\ref{fig:banana}), fish orbits with 3:2 (see Fig.\ref{fig:fish}), pretzel orbits with 4:3 (see Fig.\ref{fig:pretzel}), and so on. By plotting lines that correspond to resonances on the frequency map, we gained a first idea of the orbital families that populate our sample. For the same global resonance, for instance, 4:3, this kind of orbit can be observed in different planes (x,y), (x,z), or (y,z), that match the resonances 4:-3:0, 4:0:-3, and 0:4:-3, respectively. In addition to the previously introduced families, we also searched for other pretzel varieties: 5:4 orbits (see Fig.\ref{fig:5_4}) and 5:6 orbits (see Fig.\ref{fig:5_6}).\\
As explained Section \ref{sec:freq_determination}, we also considered the frequencies in cylindrical coordinates (see Fig.\ref{fig:freq_map_cyl}), which were only computed for $z$-tubes (because our sample contains very few $x$-tubes, we did not consider it necessary to study the frequency map in their corresponding cylindrical coordinates). There is no bisymmetry around $\Omega_{\phi}/\Omega_{R}=0$, which is another way of detecting the rotation of the NSD. More stars orbit in the positive than in the negative sense. The resonances also appear as straight lines here (horizontal lines for a resonance between the vertical oscillation frequency $\Omega_{z}$ and the radial oscillation frequency $\Omega_{R}$ and vertical lines between the azimuthal oscillation frequency $\Omega_{\phi}$ and the radial oscillation frequency $\Omega_{R}$). We discern then some resonances, $\Omega_{z}:\Omega_{R}=1:1,\ 3:4,\ 1:2$, and $\Omega_{\phi}:\Omega_{R}=1:2$. Saucer orbits (\citealt{sambhus2000};\citealt{vasiliev14:saucer}) (see an example Fig.\ref{fig:saucer}) were found via the resonance $\Omega_{z}:\Omega_{R}=1:1$ \citep{Yavetz2023}, and we therefore have an estimate of their number.\\
We call a chaotic orbit a weakly chaotic or sticky orbit when it lies near a resonance \citep{Valluri2016}. It might appear regular for some time, but will finally exhibit chaotic behaviour.
 With this method, we selected (see Fig. \ref{fig:freq_map_cart} and \ref{fig:freq_map_cyl}) all orbits close enough to the resonances in the frequency maps. More precisely, we made the selection at $\pm0.01$ (in frequency ratio), which is enough for contamination by sticky chaotic orbits. However, we limited this contamination by removing orbits from the selection of tubes, that were identified as box/chaotic orbits (without any circulation around an axis). We list in Table.\ref{table:families} the percentage for each family with and without these latter contaminants.\\


\begin{table*}
\caption{Family membership results for the automatic and visual methods}\label{table:families}
\begin{tabular}{|l|l|c|c|c|c|c|c|c|c|c|}
\cline{3-11}
    \multicolumn{2}{c|}{}  & Chaotic/Box & $x$-tube & $z$-tube$^{1}$ & Banana & Fish & Saucer & Pretzel & $5:4$ & $5:6$\\
\hline
\multirow{2}{*}{Automatic} & \textbf{Total} &  \textbf{34.0\%}  & \textbf{1.1\%} & \textbf{64.9\%} &\textbf{4.7\%}   &  \textbf{9.0\%}  & \textbf{13.6\%} & \textbf{3.5\%} & \textbf{1.5\%} & \textbf{1.6\%}\\
   & With identified chaotics & - & $2.1\%$ & - & - & $11.2\%$  & - & $5.0\%$ & $1.9\%$ & $1.8\%$\\
\hline
\multirow{3}{*}{Visual} & \textbf{Total} &  \textbf{24.7\%}  & \textbf{1.5\%} & \textbf{68.2\%} & \textbf{5.7\%}    &  \textbf{11.2\%}   & \textbf{8.7\%} & \textbf{4.5\%}  & \textbf{0.6\%} & \textbf{1.1\%}\\
   & Centrophobic & - & - & - &   $4.0\%$   & $9.6\%$  & - &  $1.8\%$ & - & -\\
   & Centrophilic & - & - & - &   $1.7\%$   & $1.6\%$  & - &  $2.6\%$ & - & -\\
\hline
\end{tabular}
\tablefoot{1: the "$z$-tube" family contains all orbits with a circulation around the z-axis and includes therefore  fish, saucer, pretzel, $5:4$, $5:6$ orbits.}
\end{table*}

\subsubsection{Visual method} \label{sec:visual}

To evaluate the validity of the automatic orbit classification, we carried out a visual classification by studying each of our 1130 orbits.
We were able to identify orbits belonging to the families presented in Section \ref{sec:automatic} that are chaotic, $x$-tube, $z$-tube, saucer, banana, fish, pretzel, $5:4$ and $5:6$ orbits. In addition to this, unlike the automatic method, the visual identification allowed us to differentiate centrophobic and centrophilic orbits of the same family (i.e. of the same resonance). Therefore, we have an estimate of the quantity of anti-banana,-fish, and -pretzel orbits. We are also able to identify orbits precisely whose own well-known resonance is lacking in the Cartesian/cylindrical frequency maps as for example saucer orbits. Because it was too difficult to visually identify box from chaotic orbits, we decided to class them in the same category for the rest of the orbital analysis as in Section.\ref{sec:automatic}.\\
As explained in Section \ref{sec:automatic}, resonant orbits have specific ratios of frequencies that correspond to their resonance. This information is clearly visible in the orbital shape. Therefore, we can visually determine the resonance of an orbit by counting how many times the orbit crossed each of the axes and obtain ratio from this. We identified the different orbital families by using this method.\\
The results of this classification show in the frequency maps of the figures \ref{fig:chaotic} to \ref{fig:5_6} that the stars belonging to the identified orbital families lie near the corresponding resonances with a larger scattering than the maximum distance from the resonance we took for the automatic method in Section \ref{sec:automatic}. However, most of the stars for each family are still located very close to the resonance. The identified chaotic/box orbits populate the frequency map in a less regular manner, as expected.\\
The location of the stars belonging to the identified orbital families in the \rmaxzmax diagram shows that they are arranged in preferential zones:\\
(i) Chaotic orbits that are scattered all over the frequency map, are only located along the observed filament in the \rmaxzmax diagram (see Fig.\ref{fig:chaotic}).\\
(ii) $x$-tubes only populate the uppermost part of the \rmaxzmax diagram, that is, they have the highest \zmax values for a wide range of \rmax (see Fig.\ref{fig:x_tube}).\\
(iii) Banana (and anti-banana) orbits are found to occupy a few wedges, but seem to be mainly present in the upper part of the \rmaxzmax diagram (see Fig.\ref{fig:banana}).\\
(iv) Fish (and anti-fish) orbits are scattered and lack a strong tendency for any wedge (see Fig.\ref{fig:fish}).\\
(v) Saucer orbits populate the middle part of the diagram and only appear to correspond to these \rmax, \zmax values (see Fig.\ref{fig:saucer}).\\
(vi) Pretzel (and anti-pretzel) orbits and their derivatives (5:4 and 5:6 orbits) are mainly located in the upper wedges (see Fig.\ref{fig:pretzel},\ref{fig:5_4},\ref{fig:5_6} respectively).\\

\begin{figure*}[ht!]
   \centering
   \includegraphics[width=0.9\textwidth]{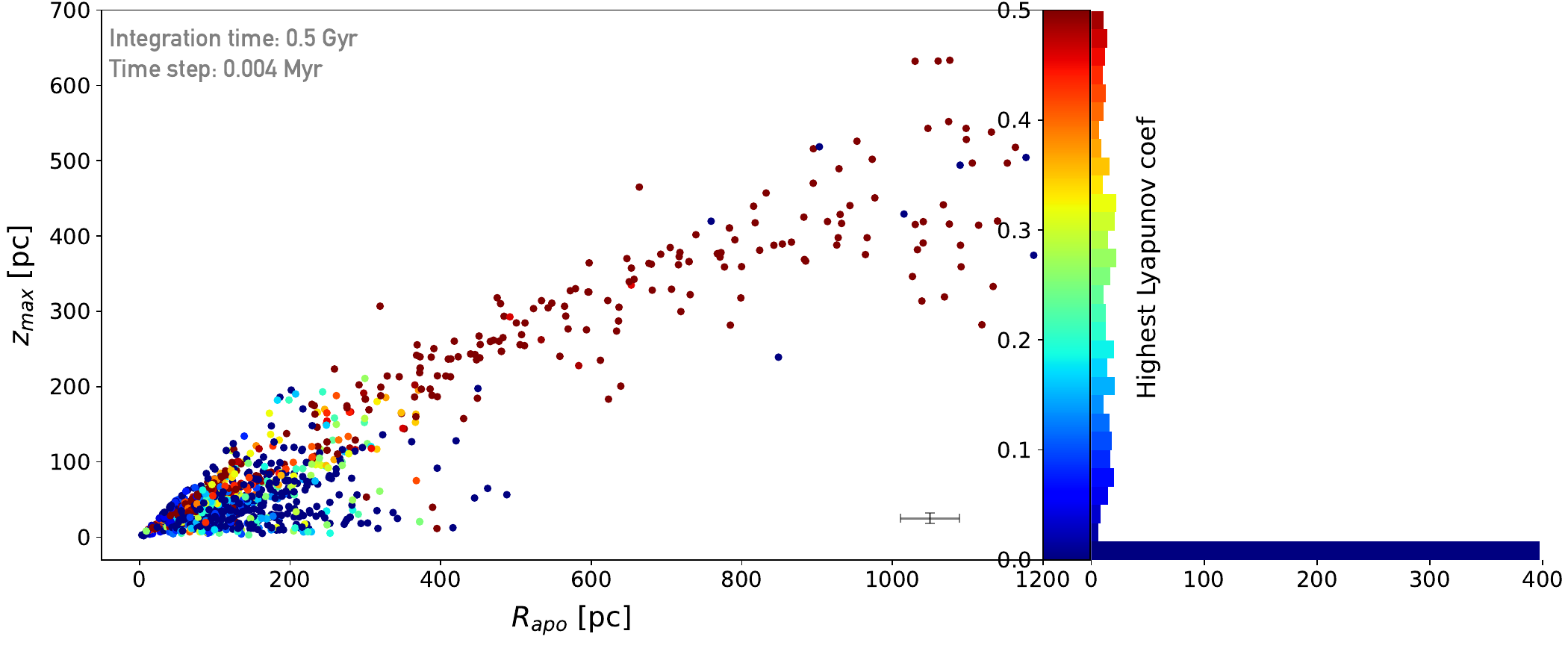}
   \includegraphics[width=0.9\textwidth]{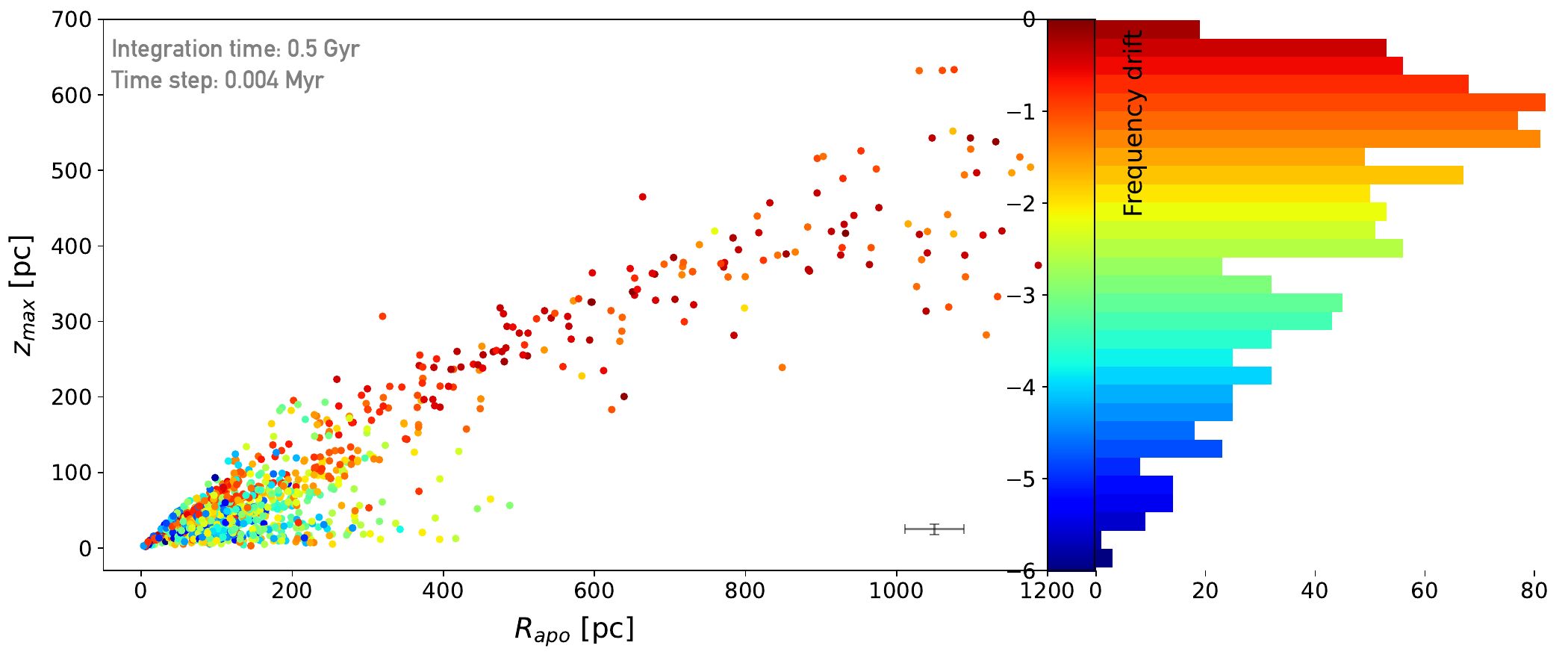}
      \caption{\textit{Upper panel:} \rmaxzmax vs. Lyapunov. \textit{Lower panel:} \rmaxzmax vs. frequency drift. Red indicates the most chaotic orbits in both methods. The typical error on both parameters arising from the propagation of the observational uncertainties and the distance uncertainty is indicated in the lower right corner of each plot.}
         \label{fig:rmax_zmax_chaos}
\end{figure*}

\begin{figure*}[ht!]
   \centering
   \includegraphics[width=0.8\textwidth]{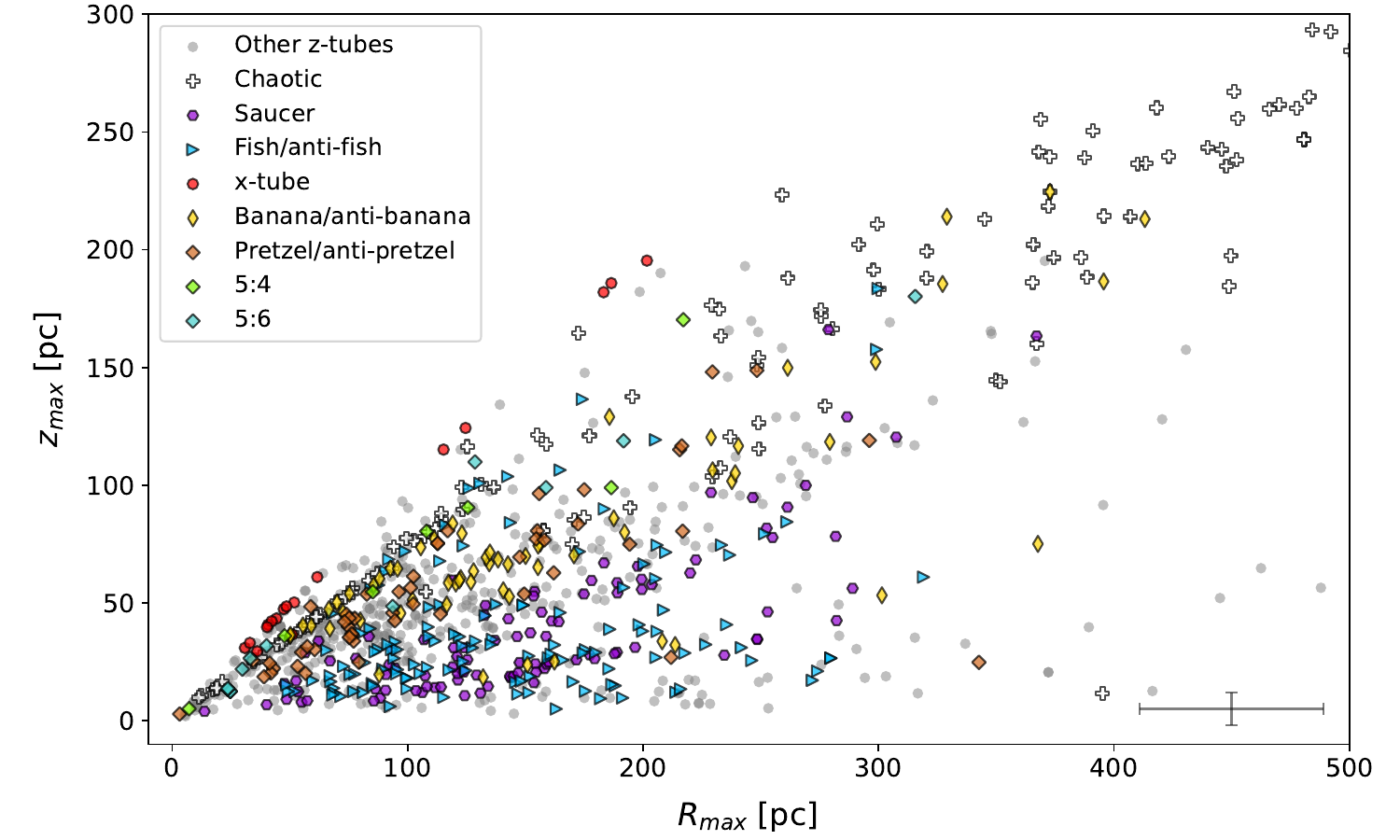}
      \caption{\rmaxzmax diagram with the different identified orbital families. Separate diagrams for each family are given in Appendix \ref{appendix:families}. The typical error on both parameters arising from the propagation of the observational uncertainties and the distance uncertainty is indicated in the lower right corner.}
         \label{fig:all_families_rmax_zmax}
\end{figure*}

\section{Discussion}\label{sec:discussion}

We made the first orbital analysis of stars located in the NSD and also using a non-axisymmetric potential. The purpose of this study therefore is to obtain a general picture of the dynamical signatures in the NSD by studying the orbital resonances.\\
Wedges in the \rmaxzmax diagram have been detected in the Galactic halo by \citet{haywood2018}, who attributed these substructures to some heating process related to the early phase of the Galactic disc. \citet{koppelman2021} investigated these features in the halo in detail using orbital frequencies and axisymmetric potentials. They demonstrated that these structures are due to resonant families and that the depletion around these resonances is related to non-integrable potentials with some indication of chaotic orbits (\citealt{Price-Whelan2016}).\\

As explained Section \ref{orbits}, we integrated orbits using four different distance estimations: a case with a fixed distance value, another case with distances chosen from a Gaussian distribution, one case with two different distances depending on the estimated position of stars along the line of sight, that is, in front or behind, and a final case where the distance of each star was derived from its position probability along the line of sight.  For all the cases, we found the similar \rmaxzmax diagrams and frequency maps, showing a very similar orbit distribution. These wedges therefore do not depend on the assumed distance inputs. This confirms that the observed substructures are indeed real and that our frequency maps are reliable. Therefore, we clearly detect wedges in the NSD, as shown in Fig.\ref{fig:rmax_zmax_ecc}, where the resonances are visible as thin straight lines.
In addition, we conducted many tests with different Galactic potentials, for instance, assuming a classical Milky Way potential with and without a rotating bar (non-axisymmetric and axisymmetric; see Fig. \ref{fig:axivsnonaxi}) with which the resonances still occur. This confirms that the conclusions of \citet{koppelman2021} are relevant in this case as well.\\
The \rmaxzmax diagrams Fig.\ref{fig:rmax_zmax_chaos} and Fig.\ref{fig:chaotic} show a "filament" at high \zmax that dives into the structure at low \rmax. It is mainly composed of chaotic orbits, regardless of the method used to identify them (highest Lyapunov exponent Sect.\ref{sec:lyapunov}, frequency drift Sect.\ref{sec:freqdrift} and visual classification Sect.\ref{sec:visual}). Because this structure is only visible when the bar potential is used in the combination of potentials and because it becomes denser with increasing integration time, we can conclude that the bar disturbs orbits to the point of making them chaotic. With an axisymmetric potential (i.e. without a bar), this filament disappears. This supports the argument of a clear bar signature of this filament. In addition, a projection of these stars in the (l,b) plane shows that these stars are mostly located in the outer parts of the NSD ($\rm |l| > 1.0^{o}$), where the contamination of the bulge/bar stars in the NSD  increases  significantly (see Fig. 10 of \citealt{Sormani2022}). Our observed filament consists of $\rm \sim 20\%$ stars of our sample, which agrees with the predicted contamination rate from the NSD models of $\rm \sim 25\%$ (\citealt{Sormani2022}).\\

The distribution of orbits in the Cartesian frequency map Fig.\ref{fig:freq_map_cart}, shows a large majority of $z$-tubes that is expected for stars forming a disc structure such as the NSD and a minority of $x$-tubes. In addition, we were able to identify a variety of families that mostly belong to $z$-tubes. We introduced in Sect.\ref{sec:results} two orbital classification methods used in this study: an automatic (see Sect.\ref{sec:automatic}) and visual method (see Sect.\ref{sec:visual}). Based on previous works about orbit classification (\citealt{valluri2010,valluri2012,Valluri2016}), we were able to search several orbital families. Therefore, the automatic and visual classifications contain the same families and we compare their number in Table.\ref{table:families}. The results of the two methods agree very well for the different identified families. However, the automatic way seems to slightly overestimate chaotic, saucer, $5:4$, and $5:6$ orbits and underestimate $z$-tube, $x$-tube, banana, fish and pretzel orbits. Even though the visual method can lead to false positives, it still remains better than the automatic one because of the visual check that allowed us to identify every orbital family without any dependence on resonance knowledge. In the era of machine learning (ML) techniques, the visual classification might in future be performed using different algorithms. \citet{ML2021} tested different ML techniques such as a random forest technique, which is a supervised decision tree algorithm, a support vector machine, or K-nearest neighbours to classify galaxies in, and around clusters according to their projected phase-space position. In order to obtain a precise classification,  a large training set is necessary on which N-body simulations can be used.   \\
Figure \ref{fig:all_families_rmax_zmax} shows that fish and saucer orbits dominate the middle and low part of the \rmaxzmax diagram, unlike banana and pretzel (and 5:4, 5:6) orbits, which cover the high part. In addition, the few $x$-tubes are located at the highest \zmax. Therefore, the results given by the visual classification confirm the link between the observed wedges in the \rmaxzmax diagram and orbital resonances, and more precisely, orbital families.\\

\begin{figure*}[ht!]
   \centering
   \includegraphics[width=1\textwidth]{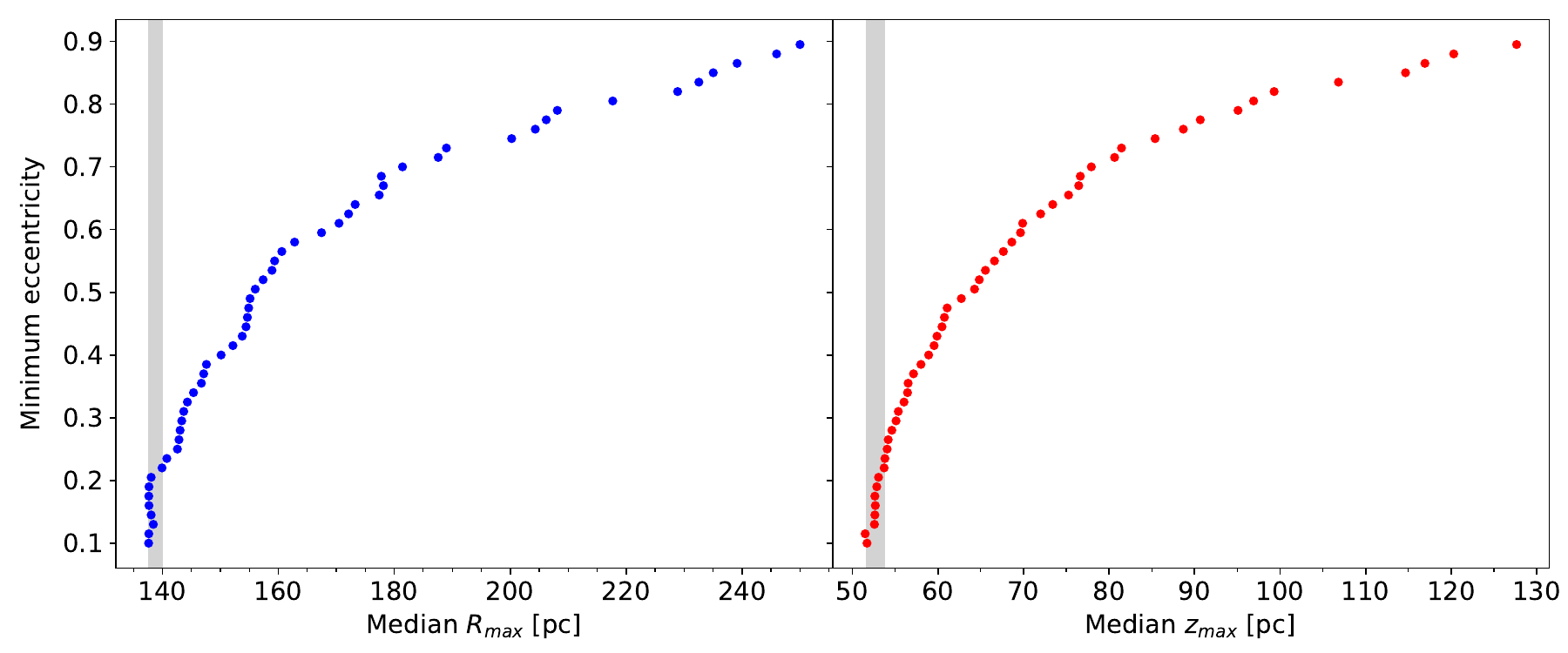}
      \caption{Median apocentric radius \rmax and maximum height \zmax for stars with a different minimum eccentricity threshold. For instance, the median \rmax for a minimum eccentricity of 0.5 means that we only computed the median of the \rmax values of stars with an eccentricity above 0.5.}
         \label{fig:nsd_dimension}
\end{figure*}

One direct application of the orbital parameter is the determination of the radial and vertical extent of the NSD. Figure \ref{fig:nsd_dimension} shows the  median apocentric radius \rmax (left panel) and \zmax for stars with a different minimum eccentricity threshold ($ecc_{min}$). For $\rm ecc_{min} < 0.25$, the truncation radius of the NSD (indicated by the grey area) is $\rm 138\pm5$ pc and the vertical extent is $53\pm1$ pc. \citet{schoenrich2015} constructed a toy model based on APOGEE kinematic data assuming a disk in which the best fit suggest a truncation radius of $\rm 150\,pc$ (see their Tab.~1). This is very consistent with the values found in this work but smaller than the radial scale  of $\rm \sim 230\,pc$ found by \citet{Launhardt2002}. This could be due to the heavy differential reddening as they used photometric data. The vertical extent is about 50\,pc in \citet{schoenrich2015}, but they did not confine this parameter from their toy model.

We stress that our sample size is limited due to our selection cuts (see Sect. 2.2), leading to statistically small sample sizes for each derived family of orbits. A first rough estimate of the completeness of our sample by using luminosity functions (see e.g. \citealt{Nogueras2023}) shows that our working sample is indeed highly incomplete ($\sim$ 15\% completeness at $\rm K_{S} \sim 11.5\,mag$).\\

Complementary  high-resolution N-body simulations of the NSD are necessary to (i) extend our study to a much larger sample, (ii) compare simulations and our observed data set in detail, and (iii) study the effect of different  Galactic potentials on the derived orbital families.\\

In the vicinity of the Galactic bar, orbits perpendicular to the bar are called $x_{2}$-type, and orbits parallel to the bar are called $x_{1}$-type orbits. $x_{2}$-type orbits are parents of the $z$-tube family, occupying the 1:-1:0 main resonance line in the frequency map (see Fig.~\ref{fig:freq_map_cart}), where about two-thirds of our sample lies. The formation of nuclear stellar discs is strongly connected to the properties of the galactic bar. Hydrodynamical simulations suggest that nuclear stellar discs form close to the inner Lindblad resonance of the main bar, where the $x_{2}$-type orbital family dominates (\citealt{Athanassoula:1992a,Athanassoula:1992b}, \citealt{Li2015}, \citealt{Sormani2018}, \citealt{Sormani2023b}). After the Galactic bar was formed, gas was funnelled along the bar towards the GC, where it settled down to form a nuclear disc. The gas then formed stars, which keep their resulting $x_{2}$ orbits, and the resulting stellar population resembles a disc.  Additional evidence comes from the spatial and kinematical overlap between the central molecular zone and the NSD (\citealt{schoenrich2015}, \citealt{Schultheis2021}) in the Milky Way, which supports this scenario that the stars in the NSD should be more metal-rich and dynamically cooler than the surrounded bar/bulge populations. This feature can  also be seen in external galaxies with NSDs (\citealt{Bittner2020}, \citealt{Gadotti2019}).\\
Planar periodic orbits $x_{1}$, $x_{2}$, and $x_{4}$ share the $1:2$ resonance between the tangential oscillation frequency ($\Omega_{\phi}$) and the radial frequency ($\Omega_{R}$). Unfortunately, they are not anticipated to be abundant in N-body simulations \citep{Valluri2016} and it is a delicate task to identify these orbits visually. This prevented us from distinguishing them properly. However, we were able to detect bifurcations from the $x_{1}$-tree \citep{athanassoula2005} with banana orbits that were called "$x_{1}v_{1}$" by \cite{skokos2002}. The presence of the latter would suggest the existence of an inner bar embedded in the NSD, as observed in others galaxies \citep{innerbarpaper}, but a more in-depth study is needed.

\section{Conclusions}

We presented a detailed orbital analysis of stars in the NSD by using orbital frequencies and a visual classification of the orbits obtained from the orbital integration. A comparison between these two methods shows very similar results in the classification of the different orbital families, such as chaotic/box, $z$-tube, $x$-tube, banana, fish, saucer, pretzel, 5:4, and 5:6 resonances.
The large majority of sources are $z$-tubes, which is expected for stars located in an NSD. We used two different methods: the Lyapunov exponent, and the frequency drift. We estimated the chaoticity of the orbits with these methods and showed by using in addition a visual classification that chaotic orbits can be identified by their high Lyapunov exponent as well as by their high frequency drift parameter.  They occupy in the \rmaxzmax the filament at high \zmax where the bar most likely causes the highly chaotic orbits.  About 20\% of our stars are contaminated by the bar/bulge population which is very close to the predicted 25\% contamination from the most recent NSD models of \citet{Sormani2022}. We emphasize that we performed here a statistical approach and that  the orbits of the individual stars can be affected by the presence of molecular clouds as well as by the disruption of their trajectory by tidal forces (e.g \citealt{Portegies2002}, \citealt{Kruijssen2014}).

We detected clear substructures in the \rmaxzmax diagram in the NSD that were identified as wedges that are related to different resonances and therefore different orbital families. 
68.2\% of our sample show $z$-tube orbits that are parented by the x2-type orbits. This is indeed expected if the formation of the NSD is coupled with the formation of the Galactic bar where x2-type orbits are the dominant population. As a follow-up work, a comparison with self-consistent models of the NSD (e.g. \citealt{Sormani2022}) is necessary for a detailed comparison between the observations and the predictions from the models (e.g. fraction of different orbital families) and to improve our orbital classification.

We used the  \rmaxzmax diagram to constrain the radial and vertical extent of the NSD  with $\rm 138\pm5$ pc and $53\pm1$ pc, respectively, which is consistent with the values found in \citet{schoenrich2015}. Due to our limited sample size, no obvious trends between the orbital families and the chemistry (e.g metallicities) is detected. Future large surveys of the NSD, such as the upcoming MOONS survey (\citealt{MOONS}) will clearly  benfit from the large number of stars to establish the strong connection between dynamics and chemistry.

\begin{acknowledgements}
We want to thank M. Valluri, A. Price-Whelan, E. Vasiliev for fruitful discussions. R. Sch\"odel acknowledges financial support from the Severo Ochoa grant CEX2021-001131- S funded by MCIN/AEI/ 10.13039/501100011033, from grant EUR2022-134031 funded by MCIN/AEI/10.13039/501100011033 and by the European Union NextGenerationEU/PRTR, and from  grant PID2022-136640NB-C21 funded by MCIU/AEI 10.13039/501100011033 and by the European Union. MCS acknowledges financial support of the Royal Society (URF\textbackslash R1\textbackslash 221118). N.N. thanks Elisa Denis and Gabriele Contursi for useful comments.\\

\textit{Softwares}: AGAMA \citep{agama}, SuperFreq \citep{superfreq}, Gala \citep{gala_software}.
\end{acknowledgements}

\bibliography{references}{} 
\bibliographystyle{aa} 

\appendix

\section{Orbital families} \label{appendix:families}

We present here typical examples of  orbits in the $x$, $y$, $z$ plane as well as the location in the \rmaxzmax diagram and in the frequency map in Cartesian coordinates. We also show in Fig.\ref{fig:freq_map_cyl_families} the location of certain families in the cylindrical frequency map.

\begin{figure*}[ht!]
   \centering
   \includegraphics[width=0.8\textwidth]{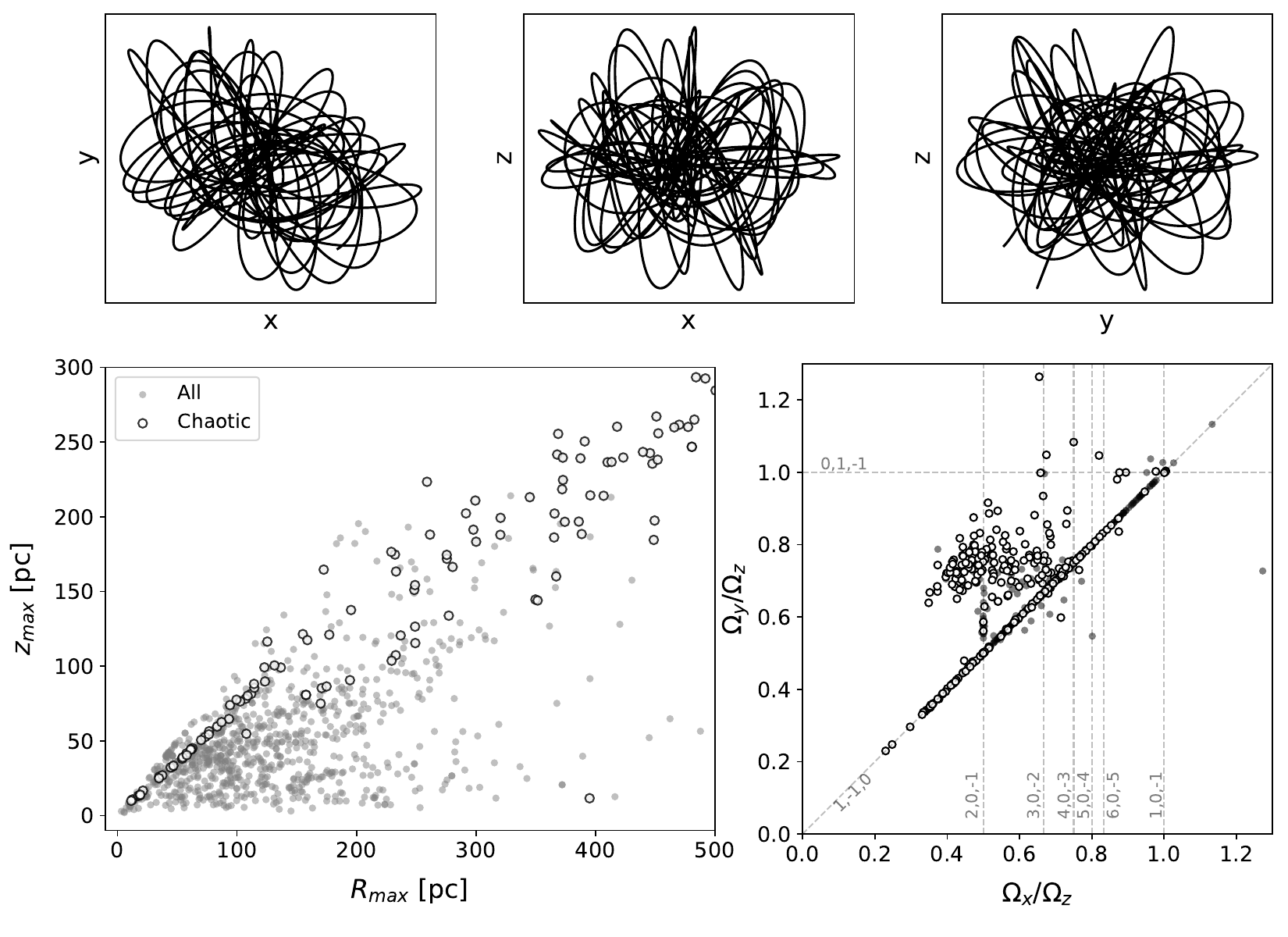}
      \caption{Example of a chaotic orbit. \textit{Upper panel:} Orbit plotted in the $(x,y)$, $(x,z)$ and $(y,z)$ planes. \textit{Lower panel:} \rmaxzmax diagram \textit{(left)} and Cartesian frequency map \textit{(right)}. The coloured markers correspond to the chaotic orbits identified with the visual method (see \ref{sec:visual}).}
         \label{fig:chaotic}
\end{figure*}

\begin{figure*}[ht!]
   \centering
   \includegraphics[width=0.8\textwidth]{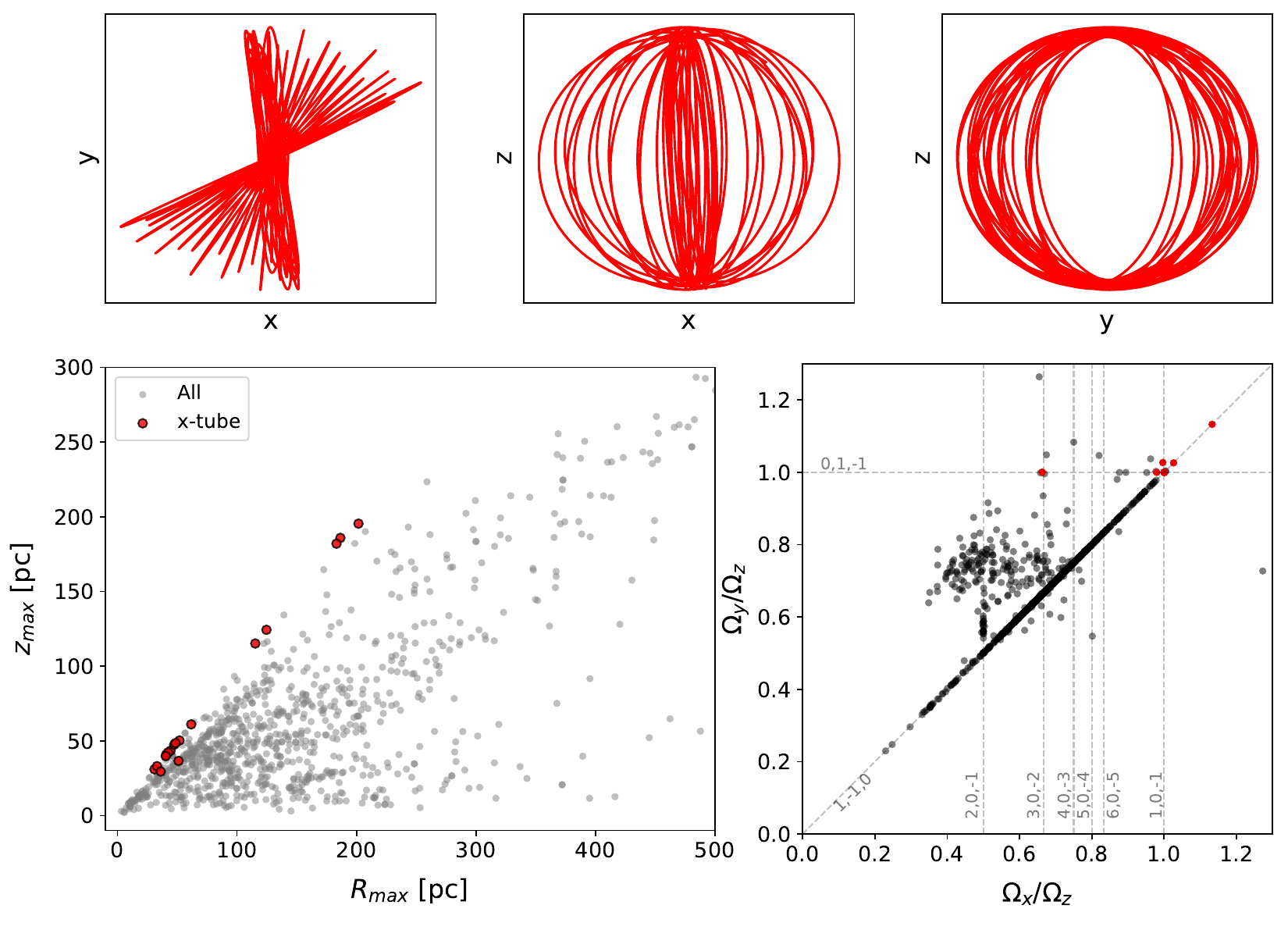}
      \caption{Example of an $x$-tube, i.e. $1:1$ resonance between $\Omega_{y}$ and $\Omega_{z}$, also called $(0:1:-1)$. \textit{Upper panel:} Orbit plotted in the $(x,y)$, $(x,z)$, and $(y,z)$ planes. \textit{Lower panel:} \rmaxzmax diagram \textit{(left)} and Cartesian frequency map \textit{(right)}. The coloured markers correspond to the $x$-tube orbits identified with the visual method (see \ref{sec:visual}).}
         \label{fig:x_tube}
\end{figure*}

\begin{figure*}[ht!]
   \centering
   \includegraphics[width=0.8\textwidth]{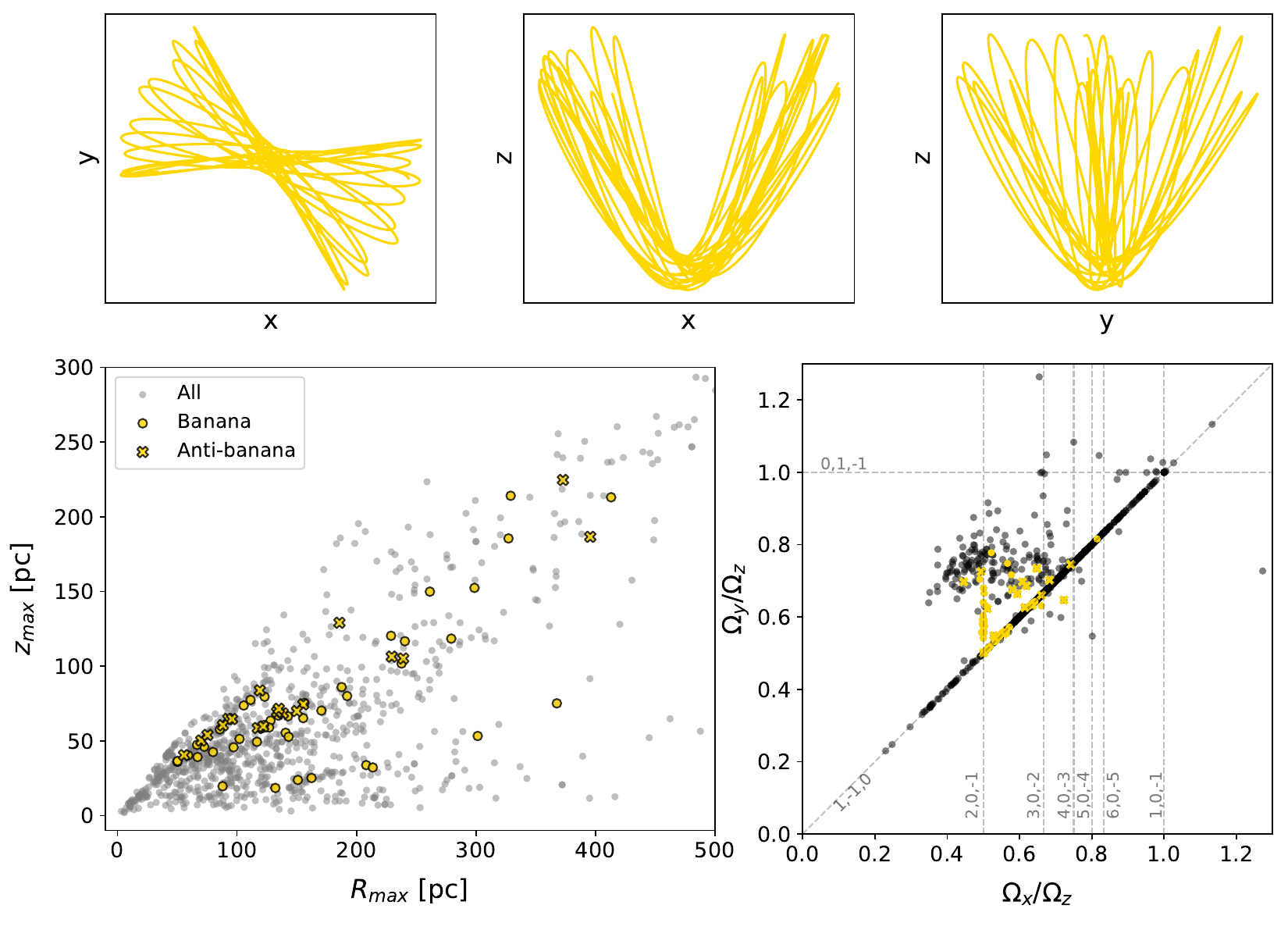}
      \caption{Example of a banana orbit (here, a $(x,z)$ banana), i.e. $2:1$ resonance between two orbital frequencies (here, $\Omega_{z}$ and $\Omega_{x}$, also called $(2:0:-1)$). \textit{Upper panel:} Orbit plotted in the $(x,y)$, $(x,z)$, and $(y,z)$ planes. \textit{Lower panel:} \rmaxzmax diagram \textit{(left)} and Cartesian frequency map \textit{(right)}. The coloured markers correspond to the banana and anti-banana orbits identified with the visual method (see \ref{sec:visual}).}
         \label{fig:banana}
\end{figure*}

\begin{figure*}[ht!]
   \centering
   \includegraphics[width=0.8\textwidth]{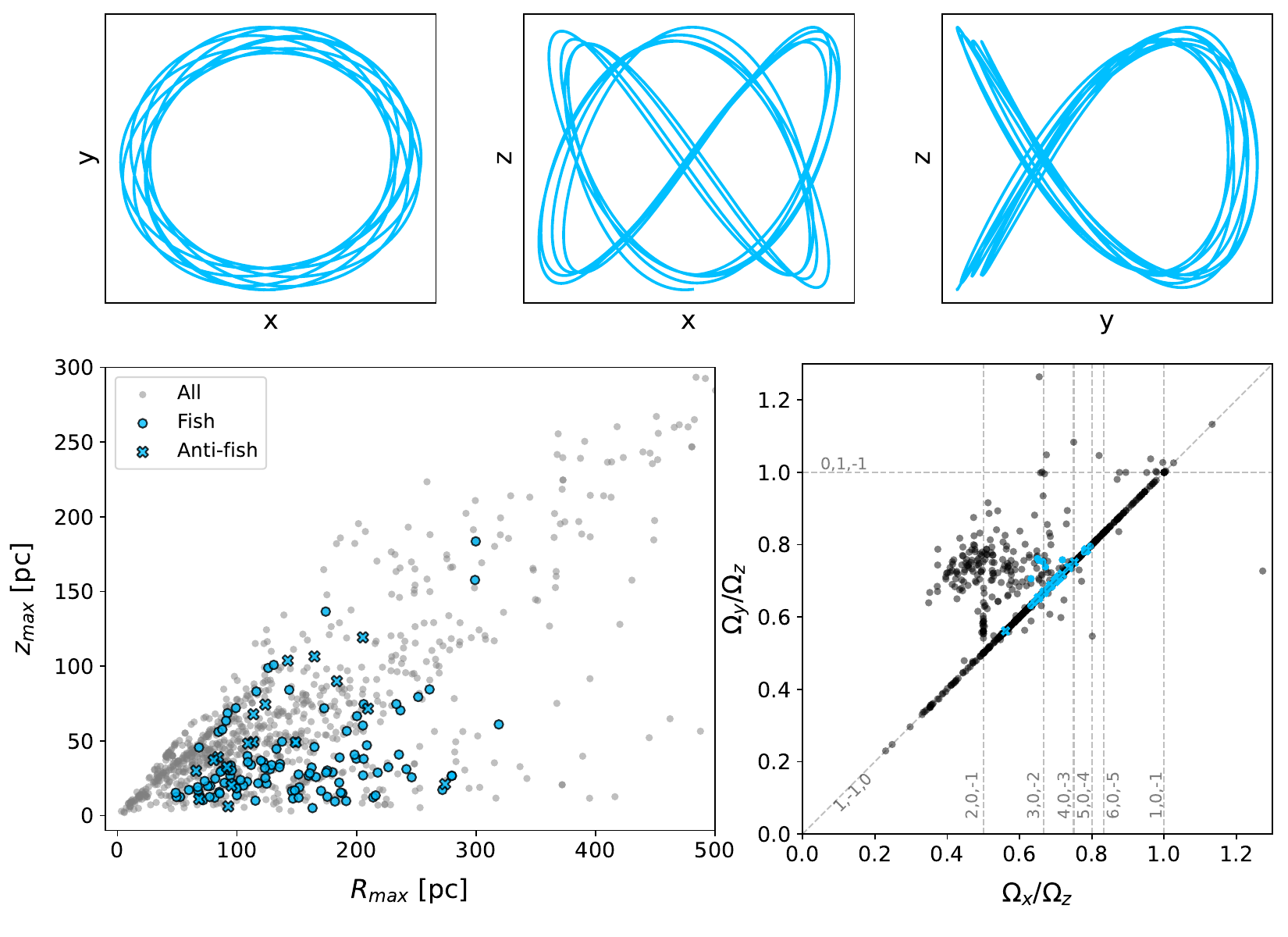}
      \caption{Example of a fish orbit (here, a $(y,z)$ fish), i.e. $3:2$ resonance between two orbital frequencies (here, $\Omega_{z}$ and $\Omega_{y}$, also called $(0:3:-2)$). \textit{Upper panel:} Orbit plotted in the $(x,y)$, $(x,z)$ and $(y,z)$ planes. \textit{Lower panel:} \rmaxzmax diagram \textit{(left)} and Cartesian frequency map \textit{(right)}. The coloured markers correspond to the fish and anti-fish orbits identified with the visual method (see \ref{sec:visual}).}
         \label{fig:fish}
\end{figure*}

\begin{figure*}[ht!]
   \centering
   \includegraphics[width=0.8\textwidth]{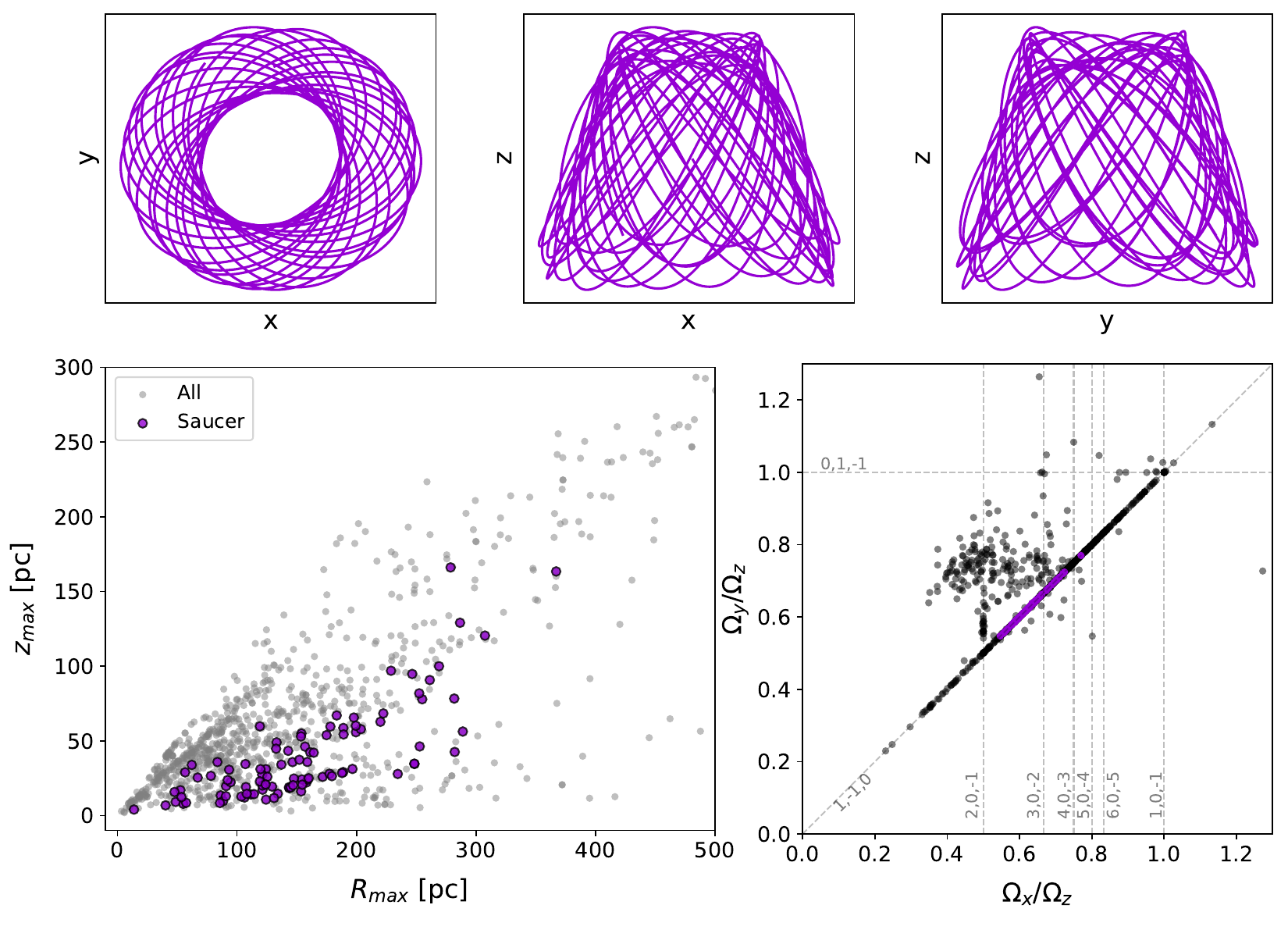}
      \caption{Example of a saucer orbit, i.e. $1:1$ resonance between $\Omega_{z}$ and $\Omega_{R}$. \textit{Upper panel:} Orbit plotted in the $(x,y)$, $(x,z)$, and $(y,z)$ planes. \textit{Lower panel:} \rmaxzmax diagram \textit{(left)} and Cartesian frequency map \textit{(right)}. The coloured markers correspond to the saucer orbits identified with the visual method (see \ref{sec:visual}).}
         \label{fig:saucer}
\end{figure*}

\begin{figure*}[ht!]
   \centering
   \includegraphics[width=0.8\textwidth]{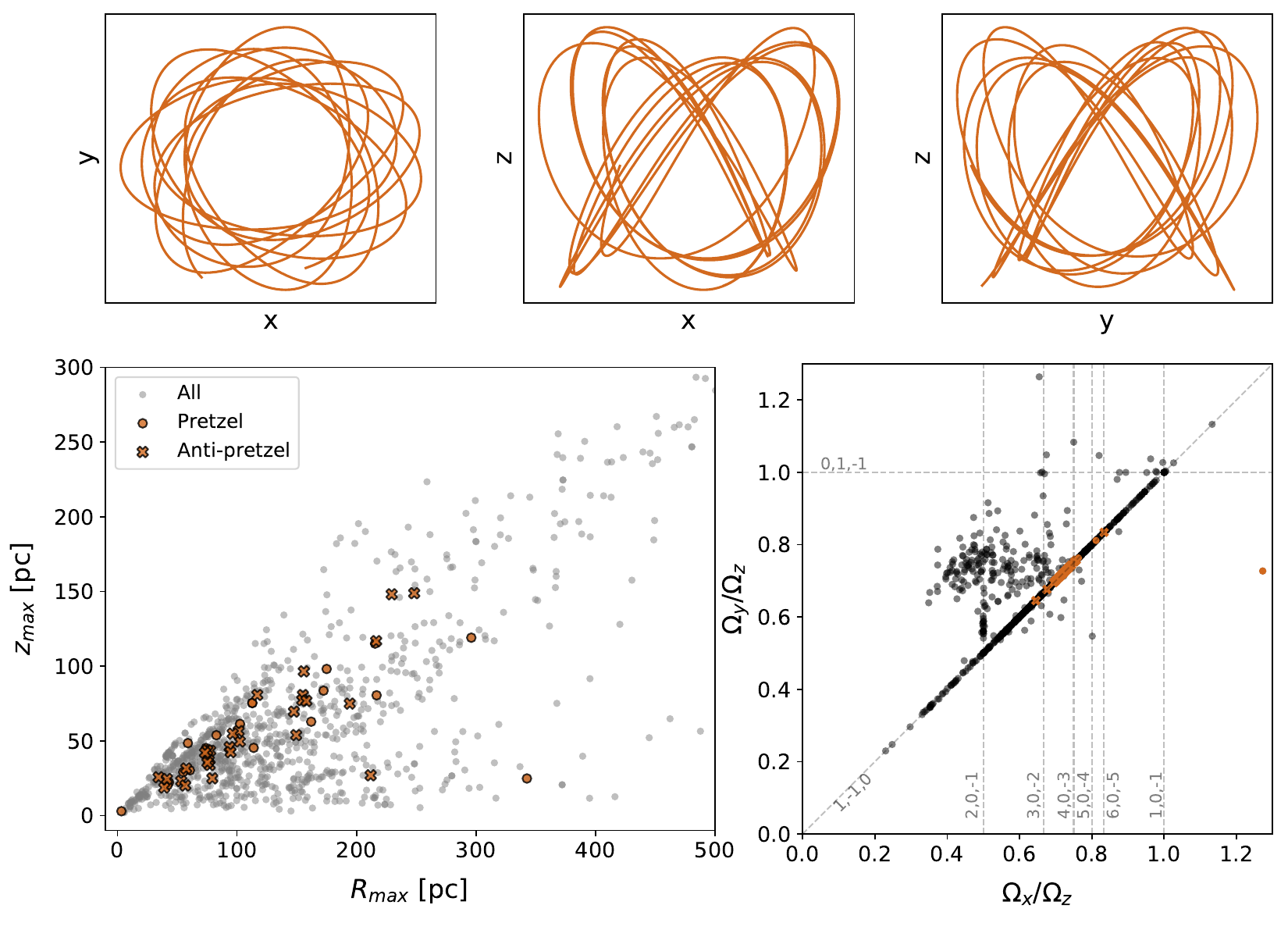}
      \caption{Example of a pretzel orbit, i.e. $4:3$ resonance between two orbital frequencies (here, $\Omega_{z}$ and $\Omega_{y}$, also called $(0:4:-3)$). \textit{Upper panel:} Orbit plotted in the $(x,y)$, $(x,z)$ and $(y,z)$ planes. \textit{Lower panel:} \rmaxzmax diagram \textit{(left)} and Cartesian frequency map \textit{(right)}. The coloured markers correspond to the pretzel and anti-pretzel orbits identified with the visual method (see \ref{sec:visual}).}
         \label{fig:pretzel}
\end{figure*}

\begin{figure*}[ht!]
   \centering
   \includegraphics[width=0.8\textwidth]{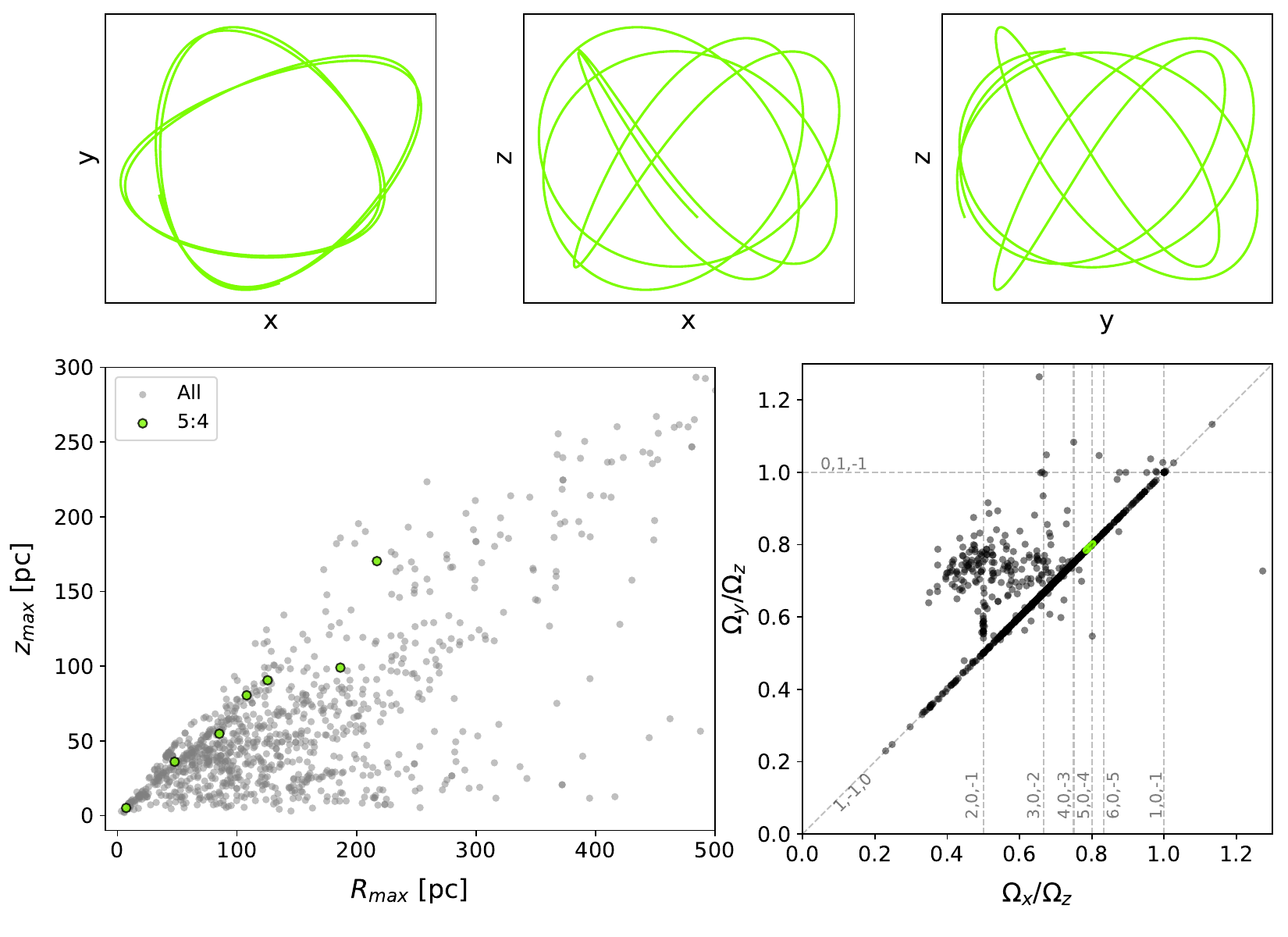}
      \caption{Example of a 5:4 orbit, i.e. $5:4$ resonance between two orbital frequencies (here $\Omega_{z}$, and $\Omega_{y}$, also called $(0:5:-4)$). \textit{Upper panel:} Orbit plotted in the $(x,y)$, $(x,z)$, and $(y,z)$ planes. \textit{Lower panel:} \rmaxzmax diagram \textit{(left)} and Cartesian frequency map \textit{(right)}. The coloured markers correspond to the 5:4 orbits identified with the visual method (see \ref{sec:visual}).}
         \label{fig:5_4}
\end{figure*}

\begin{figure*}[ht!]
   \centering
   \includegraphics[width=0.8\textwidth]{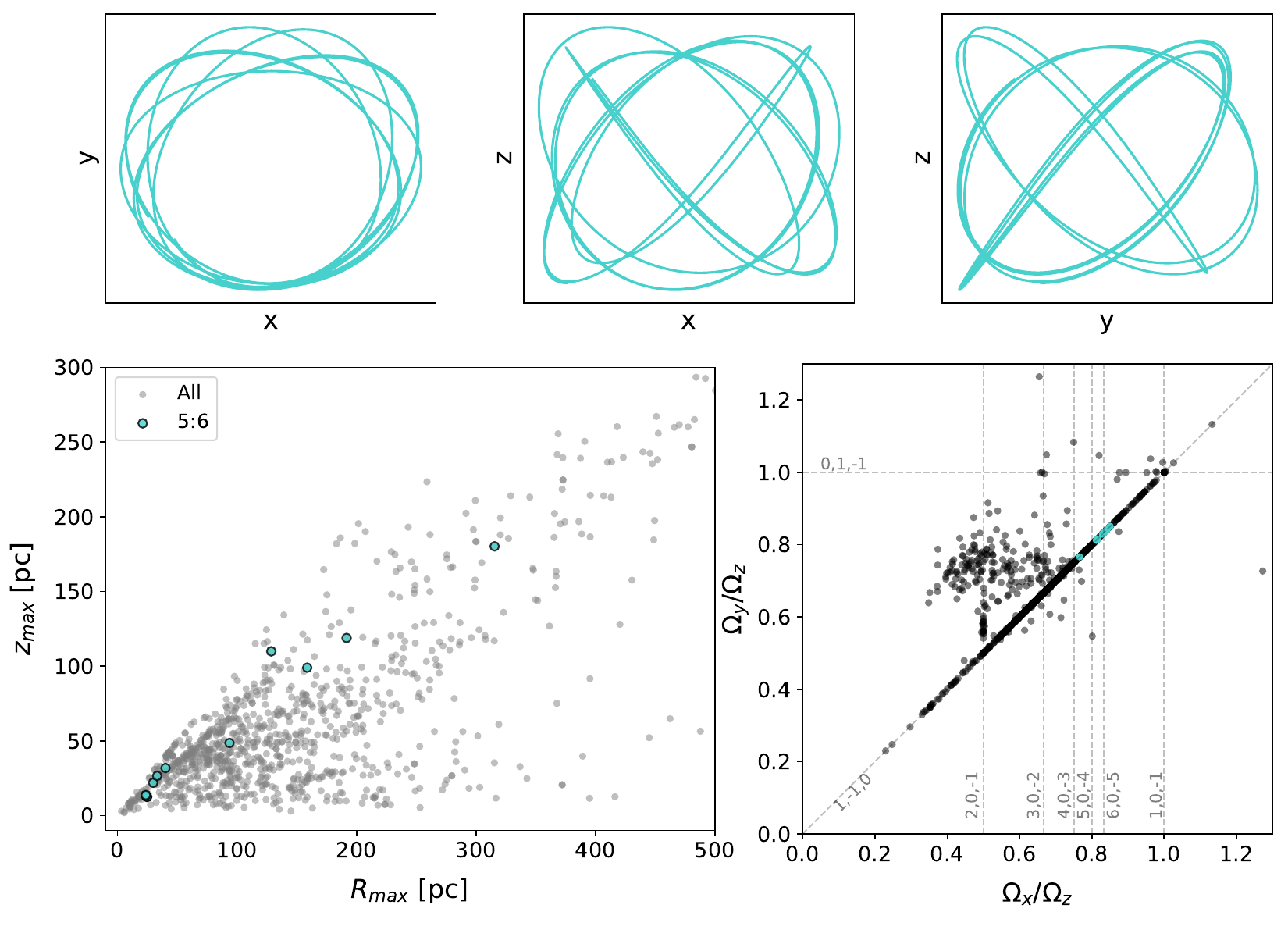}
      \caption{Example of a 5:6 orbit, i.e. $5:6$ resonance between two orbital frequencies (here, $\Omega_{y}$ and $\Omega_{z}$, also called $(0:6:-5)$). \textit{Upper panel:} Orbit plotted in the $(x,y)$, $(x,z)$, and $(y,z)$ planes. \textit{Lower panel:} \rmaxzmax diagram \textit{(left)} and Cartesian frequency map \textit{(right)}. The coloured markers correspond to the 5:6 orbits identified with the visual method (see \ref{sec:visual}).}
         \label{fig:5_6}
\end{figure*}

\begin{figure*}[ht!]
   \centering
   \includegraphics[width=0.9\textwidth]{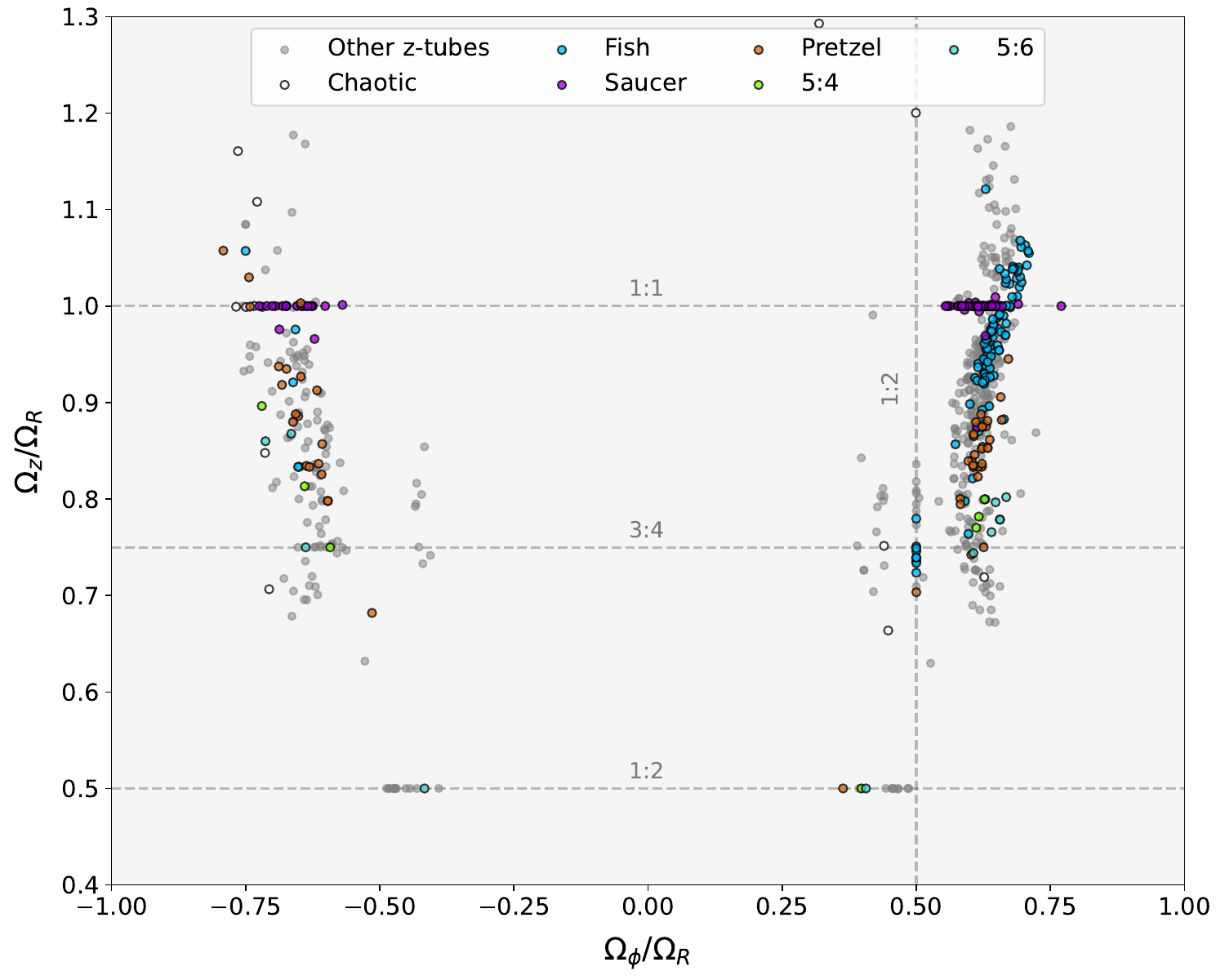}
      \caption{Frequency map in cylindrical coordinates for different families (identified with the visual method; see Section.\ref{sec:visual}) with computable cylindrical frequencies.}
         \label{fig:freq_map_cyl_families}
\end{figure*}

\section{Comparison of potentials}\label{sec:potentialscomparison}

To assess the sensitivity of the study outcomes to the selection of the mean potential, we integrated the orbits using an axisymmetric potential or slightly varied combinations of (non-axisymmetric) potentials. As detailed in section \ref{orbits}, our approach involves considering three components: the bulge/bar, the NSD, and the NSC. In the following figures (Fig. \ref{fig:axivsnonaxi}, Fig. \ref{fig:nsd_potentials}, \ref{fig:bar_potentials} and \ref{fig:bar_angle}), we compare the \rmaxzmax diagram presented in this paper (Fig. \ref{fig:rmax_zmax_ecc}) with those obtained with an axisymmetric potential (MWPotential14 from \citealt{galpy}) or by changing the NSD potential (\citealt{Sormani2022} and \citealt{Launhardt2002}), the bar potential (\citealt{Portail17}), or the bar angle.
\begin{figure*}[ht!]
   \centering
   \includegraphics[width=1\textwidth]{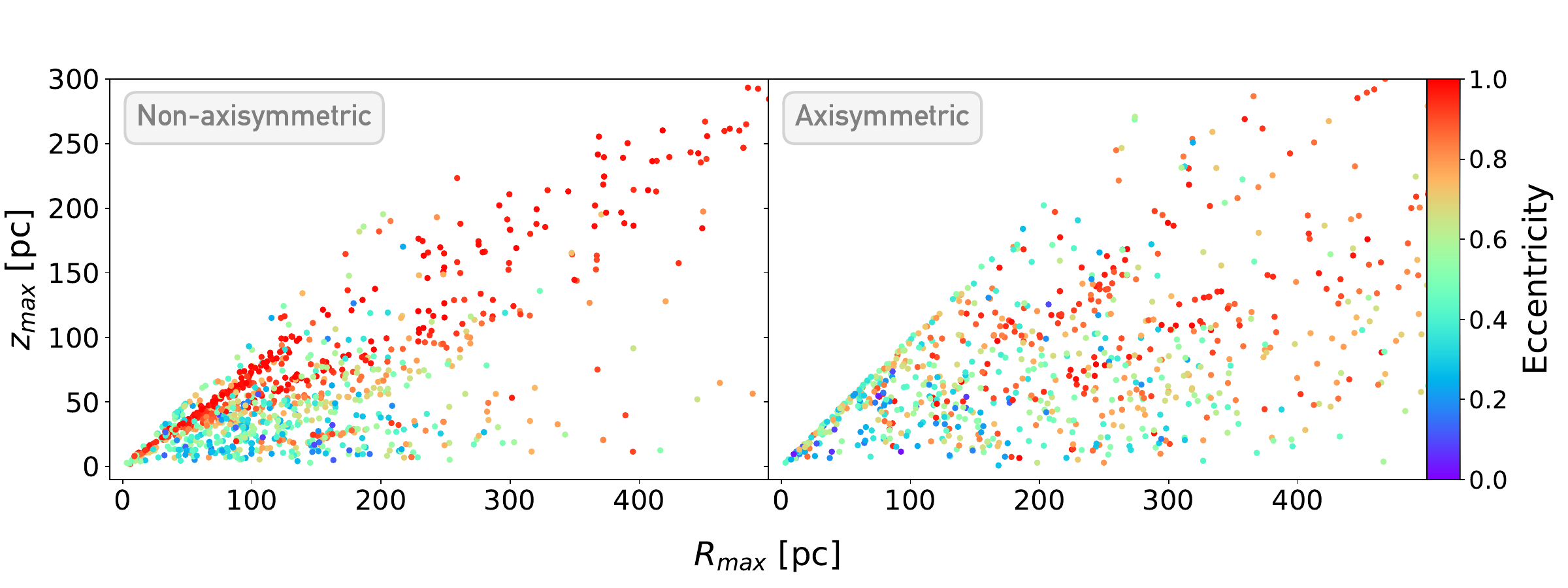}
      \caption{\rmaxzmax diagram comparison for the axisymmetric and non-axisymmetric cases. \textit{Left:} Reference diagram, with the non-axisymmetric combination of potentials presented in Section \ref{orbits}. \textit{Right:} Using the axisymmetric potential MWPotential14 from \cite{galpy}.}
         \label{fig:axivsnonaxi}
\end{figure*}

\begin{figure*}[ht!]
   \centering
   \includegraphics[scale=0.65]{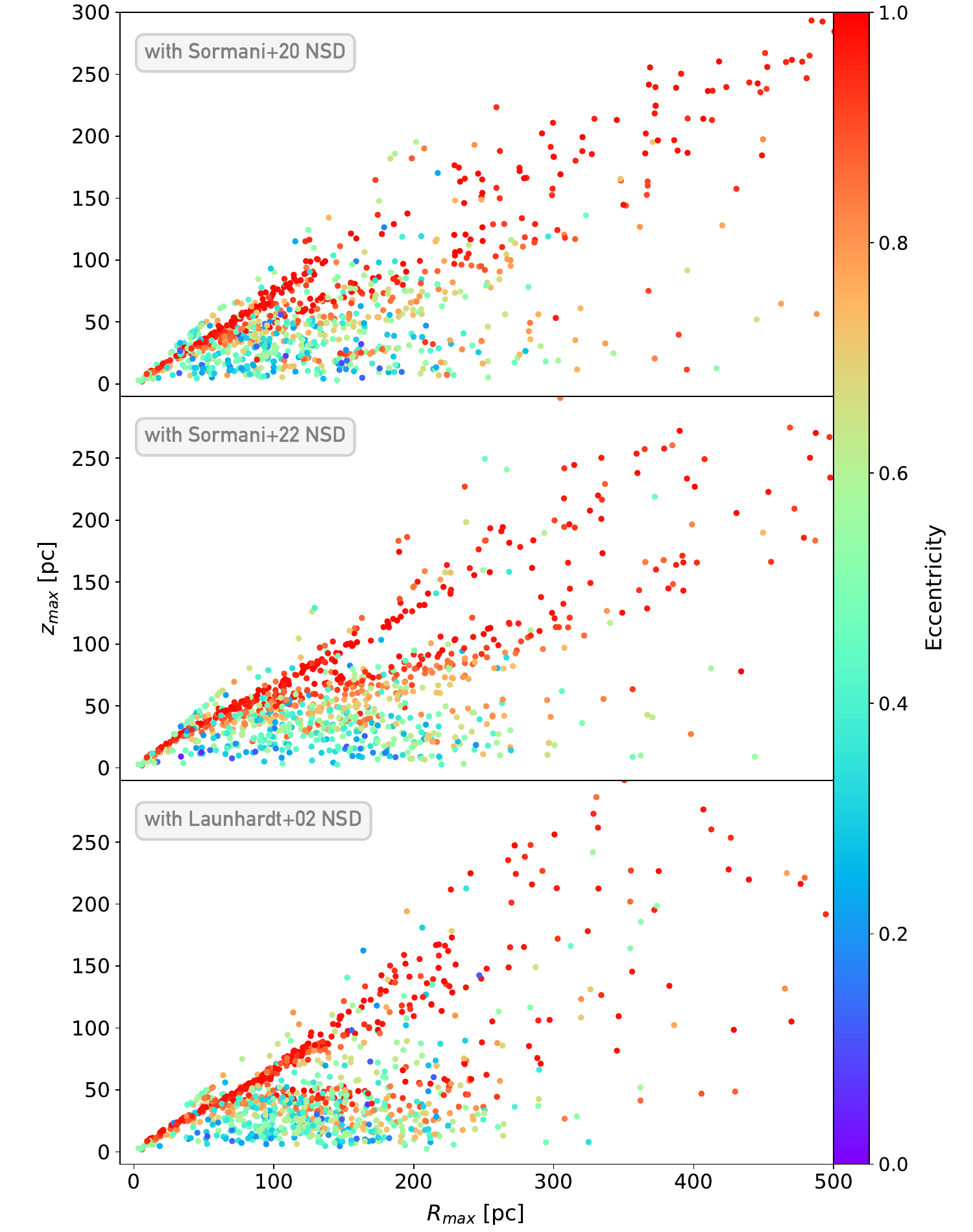}
      \caption{\rmaxzmax diagram comparison with different NSD potentials. \textit{Upper panel:} Reference diagram, with the NSD potential from \cite{Sormani2020} (model 3) used in our study. \textit{Middle panel:} Using the NSD potential from \cite{Sormani2022}. \textit{Lower panel:} Using the NSD potential from \cite{Launhardt2002}. In the three cases, the bar and the NSC potentials are not changed.}
         \label{fig:nsd_potentials}
\end{figure*}

\begin{figure*}[ht!]
   \centering
   \includegraphics[width=1\textwidth]{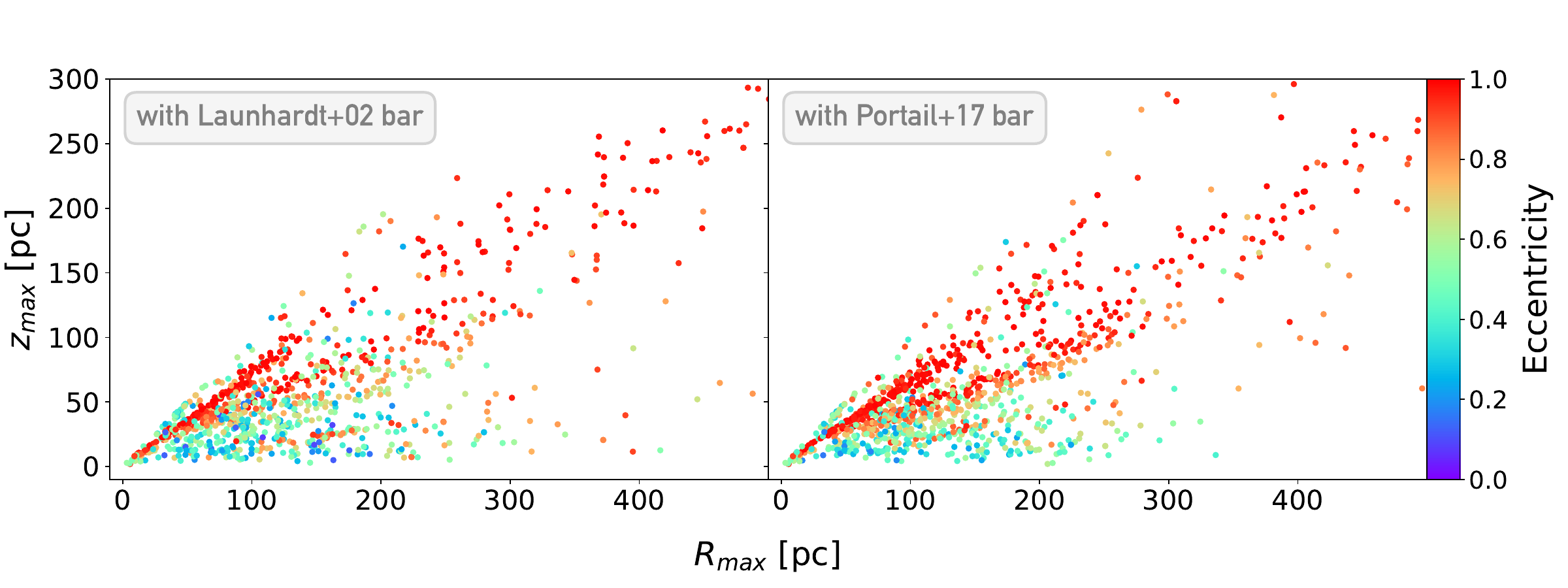}
      \caption{\rmaxzmax diagram comparison with different bar potentials. \textit{Left:} Reference diagram, with the bar potential from \cite{Launhardt2002}. \textit{Right:} Using the bar potential from \cite{Portail17}. In both cases, the NSD and NSC potentials are not changed.}
         \label{fig:bar_potentials}
\end{figure*}

\begin{figure*}[ht!]
   \centering
   \includegraphics[width=1\textwidth]{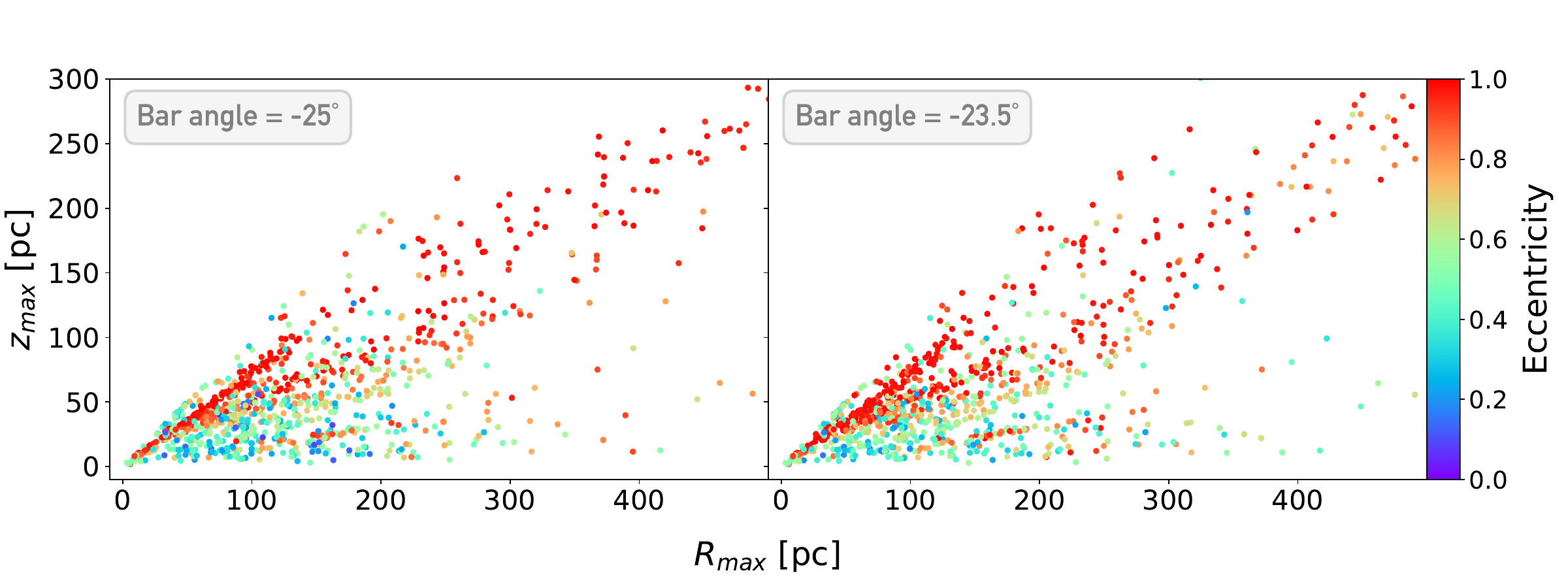}
      \caption{\rmaxzmax diagram comparison with a different bar angle with respect to the line of sight towards the GC. \textit{Left:} Reference diagram as presented in Section \ref{orbits}. \textit{Right:} Case with a $10\%$ variation of the angle.}
         \label{fig:bar_angle}
\end{figure*}

%
%

\end{document}